\documentclass[superscriptaddress,floatfix,amsmath,amssymb,pre,citesort,longbibliography]{revtex4-2}

\usepackage{hyperref}
\usepackage[usenames,dvipsnames,svgnames,table]{xcolor}
\usepackage{pgfplots,tikz}
\usepackage{array}
\usepackage{graphicx}
\usepackage{subcaption}
\usepackage{float}
\usepackage{caption}
\usepackage{bm}
\def\eg{{\it e.g.\ }}
\def\ie{{\it i.e.\ }}

\begin{document}
\title{Predator-prey plankton dynamics in turbulent wakes behind islands}

\author{Alice Jaccod}
\affiliation{Sorbonne Universit\'e, CNRS, Institut Jean Le Rond d'Alembert, F-75005 Paris, France}

\author{Stefano Berti}
\affiliation{Univ. Lille, ULR 7512 - Unit\'e de M\'ecanique de Lille Joseph Boussinesq (UML), F-59000 Lille, France}

\author{Enrico Calzavarini}
\affiliation{Univ. Lille, ULR 7512 - Unit\'e de M\'ecanique de Lille Joseph Boussinesq (UML), F-59000 Lille, France}

\author{Sergio Chibbaro}
\affiliation{Sorbonne Universit\'e, CNRS, Institut Jean Le Rond d'Alembert, F-75005 Paris, France}
		
\begin{abstract}
\label{sec:ab}
Plankton constitutes the productive base of 
aquatic ecosystems and plays an 
essential role in the global carbon 
cycle.  
The impact of 
hydrodynamic conditions on the biological activity of plankton species 
can manifest in a variety of different ways and the understanding of the basic effects  
due to multiscale complex flows still appears incomplete. 
Here, we consider a predator–prey model of plankton dynamics in the presence of a 
turbulent flow past an idealized island, 
to investigate the conditions under which an algal bloom is observed, and the resulting patchiness of plankton distributions. 
By means of direct numerical simulations, we explore the 
role of the turbulent intensity 
and of the island shape. In particular, we focus on population variance spectra, and on their relation 
with the statistical features of the turbulent flow, as well as on the correlation between the spatial distribution 
of the planktonic species and velocity field persistent structures. We find that both 
the average biomass
and local plankton dynamics critically depend on the relation between advective and biological time scales, 
confirming and extending previous results obtained in simpler flow settings. 
We further provide evidence that, in the present system, due to local flow action plankton accumulates in localized filamentary regions.
Small turbulent scales are found to impact the statistics of plankton density fields at very fine scales, 
but the main global features of the population dynamics only weakly depend on the Reynolds number and are also found to be remarkably independent of the geometrical details of the obstacle.
\end{abstract}
\maketitle


\section{Introduction}	
\label{sec:intro}
Planktonic populations are key to aquatic ecosystems, as they form the base of marine and lacustrine food webs~\cite{ML2005}, 
and play a central role in the climate, by taking part in the global carbon budget~\cite{WF2011}. 
Plankton blooms, however, can also have major negative environmental and societal impacts, 
when involving harmful algae~\cite{Sellner_etal_2003,KM2008,GF2020}. 
The modeling of plankton dynamics is therefore relevant to different studies, both from a 
fundamental and an applied point of view.

Among the physical-biological interactions controlling plankton blooms a prominent role is played by fluid transport.  
Understanding the role of laminar and turbulent flows on this phenomenon, both on the vertical~\cite{denman1995biological,Huisman_etal_2002,Lindemann_etal_2017},  
and on the horizontal~\cite{abra,martin2002,martin2003phytoplankton,Reigada_etal_2003,lo,GR2008,GF2020} 
has attracted considerable interest. 
The complexity of the problem arises from the different processes acting on a very broad range of spatial and 
temporal scales~\cite{Levy2008,McGillicuddy_2016,Levy_etal_2018,Zhang_etal_2019}. 
Furthermore, the interplay between the fluid and biological dynamics is often subtle, making  
the prediction of the conditions for blooming non-trivial even in relatively 
simple theoretical models~\cite{hl,Lindemann_etal_2017,GF2020}.

A distinctive feature characterizing plankton populations at the ocean surface is their patchiness, 
meaning their heterogeneous spatial distribution, due to lateral stirring and mixing, 
as revealed by satellite and ship-transect measurements of chlorophyll concentration 
(an indicator of the local phytoplankton biomass)~\cite{DP1976,Smith_etal_1988,MS2002,LK2004,franks}. 
Several efforts have been devoted to explain and numerically reproduce the statistical features, 
such as spectra, of plankton density fields.  
Using dimensional arguments, some theoretical predictions have been obtained, suggesting that the spectra 
of biological species in two-dimensional (2D) turbulent flows should be flatter than that of the velocity field, 
particularly for interacting species~\cite{DP1976,powell}. 
In numerical simulations, relying on an idealized model, the role of turbulent advection in the generation 
of patchiness was first put in evidence in~\cite{abra}, where some differences between phytoplankton and 
zooplankton were also observed, due to the typical biological time scale of the latter.  
The dominance of physical processes in structuring the spatial variability 
of plankton distributions was also reported in more realistic numerical studies (see, {\it e.g.},~\cite{LK2004}). 
The picture emerging from previous experimental, theoretical and numerical investigations, however, questions 
the universality of spectral slopes, pointing to large variability with respect to the physical and biological 
processes considered.

Turbulent flows of environmental interest, and particularly oceanic ones in the submesoscale and mesoscale ranges 
(horizontal size of $O(1-10)$~km and $O(10-100)$~km, respectively) are also characterized by the presence of coherent 
structures, in the form of eddies and filaments. Such structures have an important impact on biological dynamics, as 
they shape the spatial distribution of the different species~\cite{pasquero2004coherent}. 
Their effects on productivity depend on multiple factors. Based on kinematic-flow numerical simulations, 
and including nutrient dynamics, it has been argued that the confinement of plankton for sufficiently long time inside large eddies 
promotes biological growth~\cite{Sandulescu_etal2007,Sandulescu_etal2008}. 
Intense mesoscale stirring, on the other hand, can also reduce productivity in coastal upwelling systems, 
as shown using the same nutrient-phytoplankton-zooplankton (NPZ) model and flow fields from both satellite data 
and a regional model in the Benguela area~\cite{Carrasco_etal_2014}. This apparently counter-intuitive result 
confirms previous remote-sensing observations~\cite{rossi2008comparative,rossi2009surface} 
and is found to be due to relevant off-shore advection, through an analysis of the correlation between 
spatial features in plankton density fields and Lagrangian coherent structures.\\
Several important ideas to investigate the basic mechanisms underlying the effects of fluid motions 
on biological dynamics, as those mentioned above, were put forward in studies examining chemical and biological reactions 
in the presence of chaotic advection, in both closed and open flows~\cite{toroczkai1998advection,neufeld,neu,lo,NLHP2002}. 
In particular, these works highlighted the role of a special flow structure, the chaotic saddle, forming in open flows 
and capable of entraining fluid parcels for long time intervals~\cite{hl,neufeld,NLHP2002,Sandulescu_etal2007,Sandulescu_etal2008,GF2020}. 
Due to this property, when such flows are coupled to excitable biological dynamics~\cite{tru,truscott1994equilibria}, 
sustained blooms can take place~\cite{hl}. As both fluid transport across the region of interest and biological 
growth are transient phenomena in that case, such a feature provides, in our opinion, an effective illustration 
of the non-trivial interplay between fluid and reactive dynamics. 
The above studies, however, considered kinematic velocity fields, namely specified by a prescribed stream function. 
While such a simplified approach allows the description of some of the main flow features and offers reduced 
computational cost, it cannot account for complex flow dynamics, involving a whole range of temporal and spatial scales. 
Furthermore, it cannot be easily generalized to different geometries and boundary conditions.

In this work we explore the dynamics of a predator-prey model of plankton blooms (PZ)~\cite{tru}, displaying excitability, 
in turbulent flows occurring in the wake of an obstacle. 
Our flow configuration shares some similarities with the one used 
to study plankton dynamics in the Canary region~\cite{sandu2006, Sandulescu_etal2007, Sandulescu_etal2008,GF2020} 
using a kinematic flow (and NPZ model), but we take a more general perspective. 
In our setup, the obstacle mimics a generic island, rather than a specific one, 
and it could also equally represent another obstruction (\eg a man-made construction) in a current. 
More importantly, our interest is mainly centered on the possible effects due the spatiotemporal complexity of 
dynamical turbulent flows. 
In particular, we aim at identifying the minimal flow ingredients needed to sustain a bloom, and at characterizing how the latter 
could be affected by multi-scale fluid properties. 
While investigating an idealised system, we avoid any bias possibly coming from the modeling of the small scales of the flow.
Furthermore, we have chosen the PZ model to leave apart possible effects linked to 
nutrient heterogeneities.\\
For this purpose we revisit some of the theoretical predictions and the numerical results 
obtained in simplified settings~\cite{hl,lo,NLHP2002}, to test their robustness against genuine multi-scale flows. 
We then focus on the conditions for the occurrence of blooms, 
and on their intensity in terms of global biomass produced, in progressively more turbulent two-dimensional  flows. 
By means of extensive fully-resolved numerical simulations, we further investigate 
the statistical properties of plankton patchiness, quantifying variance spectra, and analyse the correlations 
between the spatial organisation of prey (phytoplankton) and predator (zooplankton) populations with flow structures. 
Since it is not possible to perform resolved simulations of realistic configurations, we evaluate the impact of varying the size of the system.
Finally, we consider the effect of changing the obstacle shape and assess the role of the roughness of its boundary, 
which had not been examined before, in spite of its relevance in realistic situations.

This article is organized as follows. In Sec.~\ref{sec: model} we introduce the mathematical framework of the problem, 
recalling the main dynamical features of the biological model adopted, and describing its coupling with the 
equations governing hydrodynamics. The numerical setup, as well as the flow configuration and the main parameters used, 
are illustrated in Sec.~\ref{sec:numerics}. 
We present the results of our numerical study in Sec.~\ref{sec:results}, particularly discussing the 
impacts on the biological dynamics of the different turbulence regimes, of the  obstacle shape and of 
under-resolving the velocity field. 
Finally, discussions and conclusions are presented in Sec.~\ref{sec:conclu}.

\section{Mathematical formulation}
\label{sec: model}

We investigate the growth dynamics of two planktonic species, the phytoplankton and the zooplankton, 
living in a fluid environment localized around islands, 
characterized by 
predator-prey interactions. 
Their spatio-temporal evolution can be conveniently described using coupled advection-reaction-diffusion equations.\\
As for the reaction kinetics, a simple model accounting in an effective way for bloom dynamics was proposed by~\cite{tru}  
based on the properties of excitable media~\cite{mu},~\cite{gri}. This model, also known as PZ (for phytoplankton-zooplankton) model, provided useful 
to reproduce the main dynamical features of red tides. Its two basic ingredients are the trigger mechanism, represented by the interaction 
between the phytoplankton growth rate and the grazing rate of zooplankton, which gives rise to the prey population 
outbreak, and the refractory mechanism, represented by the growth of zooplankton, which causes the system return to the initial equilibrium state.\\
In well-mixed conditions, calling $P=P(t)$ and $Z=Z(t)$ the population densities of phytoplankton and zooplankton, respectively, 
the evolution equations read: 
\begin{subequations}
\begin{align}
&\frac{dP}{dt} = rP\left(1-\frac{P}{K}\right) - R_m Z \frac{P^2}{P^2 + \kappa^2}, \label{P_ode}\\
&\frac{dZ}{dt} = \gamma R_m Z \frac{P^2}{P^2 + \kappa^2} - \mu Z. \label{Z_ode} 
\end{align}
\end{subequations}
The term $rP(1-P/K)$ represents the gross rate of production of phytoplankton, called primary production $PP$, and is expressed 
by a logistic growth function, with a maximum specific growth rate $r$ and a carrying capacity $K$. 
Predation of phytoplankton is represented by a Holling Type-III function~\cite{holl}, where $R_m$ is the maximum specific predation rate 
and $\kappa$ determines how quickly that maximum is attained as the prey population density increases.
The rate of zooplankton production is controlled by the population density of phytoplankton, with $\gamma$ representing the ratio of biomass 
consumed to biomass of new herbivores produced. The rate of zooplankton removal, by natural death and predation from higher organisms, is called $\mu$.  
To display excitability, the PZ model must have at least two different time scales: to initiate an outbreak, the phytoplankton growth rate must be 
larger than the predation rate by the zooplankton population.

To consider the previous reactive dynamics in a fluid flow, it is necessary to specify the evolution equation for the velocity field 
and the coupling of the latter with the population densities, $P(\bm{x},t)$ and $Z(\bm{x},t)$. 
It is useful to formulate the complete model in non-dimensional variables. For this purpose, we introduce a characteristic length $l_0$ 
and a typical velocity $u_0$, from which the typical time is $t_0 = l_0/u_0$.
We then consider an incompressible two-dimensional (2D) flow defined on a square domain of side $L$, in the presence of  
a circular obstacle (of average radius $l_0$) representing an island, which is the solution of the Navier-Stokes equation with the appropriate 
boundary conditions (see Sec.~\ref{sec:numerics}). The non-dimensional form of the latter equation and of the incompressibility condition is:
 \begin{subequations}
\begin{align}
&\partial_t \bm{u} + (\bm{u} \cdot \bm{\nabla}) \bm{u} = -\bm{\nabla} p + \frac{1}{Re} \bm{\nabla}^2 \bm{u} \label{ansa}\\
& \bm{\nabla} \cdot  \bm{u}  = 0 
\label{ansb}
\end{align}
\end{subequations}
where $\bm{u}(\boldsymbol{x},t)$ is the dimensionless fluid velocity field, $p$ is pressure and $Re = u_0d/\nu$ the Reynolds number based on the island diameter $d = 2l_0$, with $\nu$ the viscosity coefficient. 

As we have neglected any feedback effects of the planktonic species on the velocity field in Eq.~(\ref{ansa}), as it is reasonable considering 
their weak values in realistic conditions, the coupling between the biological and fluid dynamics, is realised only by advection. 
This implies that the time derivatives in Eqs.~(\ref{P_ode}-\ref{Z_ode}) now need to be interpreted as material derivatives. 
Note that, in the following we will also add a diffusivity term to both equations, with a diffusion coefficient $D$ 
(equal for the two species),  
possibly due to swimming behaviour. 
Proceeding as before and further normalizing the population densities with the carrying capacity $K$, we obtain:
\begin{subequations}
\begin{align}
& \partial_t P +  \bm{u} \cdot \bm{\nabla} P - \frac{1}{Re Sc}\bm{\nabla}^2 P = \epsilon \left(\beta P\left(1-P\right) - \delta Z \frac{P^2}{P^2 + \chi^2}\right),\label{apza}\\
& \partial_t Z +  \bm{u} \cdot \bm{\nabla} Z - \frac{1}{Re Sc}\bm{\nabla}^2 Z =  
\epsilon \gamma Z \left( \delta \frac{P^2}{P^2 + \chi^2} - \lambda \right), \label{apzb}
\end{align} 
\end{subequations}
where $Sc = \nu/D$ is the Schmidt number, $\beta = rl_0/u_0$, $\delta = R_m l_0/u_0$, $\chi = \kappa/K$, $\lambda = \mu l_0/(u_0 \gamma)$.
In addition, we have introduced the parameter $\epsilon$ in front of the reaction terms, 
which allows to perform a parametric study in a simple way, by artificially changing the ratio between the advective time scale 
and the biological activity one. In the following we will set it to $1$, unless explicitly stated.
The dynamics resulting from Eqs.~(\ref{ansa}-\ref{ansb}) and~(\ref{apza}-\ref{apzb}) 
are generally highly nontrivial, which severely limits the possibility to perform analytical calculations. 
Some results concerning the transport and mixing of planktonic species have been, nevertheless, previously obtained in simpler configurations, using 
the tools of dynamical systems theory, and will serve us as a guide for our analysis ~\cite{chaos,lo,fer,neu}. 
While the analytical resolution of the full dynamics is not possible, it is instructive to recall some important results 
concerning the reactive dynamics in the absence of flow~\cite{tru}. Indeed, from Eqs.~(\ref{P_ode}-\ref{Z_ode}), one can obtain the fixed points 
of the PZ model: $(P_1,Z_1)=(0,0)$, $(P_2,Z_2)=(1,0)$ and $(P_3,Z_3)=(P_{eq}, Z_{eq})$ where $P_{eq}=  \chi\sqrt{\lambda/(\delta-\lambda)}$ and $Z_{eq}= \beta(1-P_{eq})(P_{eq}^2+\chi^2)/(P_{eq}\delta)$, expressed in non-dimensional variables.
The first one represents the extinction of both species, $(P_2,Z_2)$ gives the equilibrium value of $P$ in the absence of $Z$ and $(P_3, Z_3)$ 
is the stable pre-outbreak state of species coexistence. From stability analysis, it emerges that $(P_1,Z_1)$ and $(P_2,Z_2)$ 
are saddle points,
while $(P_3,Z_3)$ is a stable equilibrium point when appropriate parameter values are used (see~\cite{tru} and Sec.~\ref{sec:numerics}). 

\section{Numerical setup}
\label{sec:numerics}

The flow field is assumed to be initially uniform and unidirectional, $\bm{u}=(u_0,0)$. 
In the following we will refer to such a streamwise direction as the longitudinal (or $x$) direction, 
and to the cross-stream one as the transversal (or $y$) direction.
The reference dimensional and non-dimensional values we adopted for the biological model ~\cite{hl} 
are reported in Table~\ref{tab1}.

\begin{table}[ht]
\begin{center}
$
\begin{array}{|c|c|c|}
\hline
\text{Parameter} & \text{Value} & \text{Dimensionless value}\\
\hline
K                 & 108 \,\mu \mathrm{g~N~l}^{-1}    &  1\\
r \, (\beta)         & 0.3 \, \mathrm{day}^{-1}       &  4.285\\
R_m \, (\delta)       & 0.7 \, \mathrm{day}^{-1}       &  10\\
\alpha \, (\chi)      & 5.7 \, \mu \mathrm{g~N~l}^{-1}   &  0.053\\
\mu \, (\lambda)      & 0.0024 \, \mathrm{day}^{-1}     &  3.428\\
\gamma            & 0.01                           &0.01\\
\hline
\end{array}
$
\end{center}
\caption{Parameters used in the biological model. The symbols adopted for the non-dimensional quantities appear in parentheses 
in the first column. The values are consistent with typical oceanic ones.}\label{tab1}
\end{table}
  
Once the biological parameters fixed, the control parameters are the Reynolds and Schmidt 
numbers.
Using for both of them the realistic values for oceanic conditions is beyond the capabilities of current direct numerical simulations (DNS), so that moderate $Re$ numbers have to be chosen. 
Yet we have investigated different Reynolds-number flows, to analyze the possible impact of this choice.
Based on the effective diffusivity of swimming algae~\cite{Polin_etal2009,Garcia_etal2011,Brun_etal2019}, we appraise the $Sc$ number in the range $\approx(10^2-10^3)$. 
Here, due to numerical constraints, we fix $Sc=10^2$. From this parameter, the smallest relevant scale, \ie the Batchelor scale, is
$\ell_B = \ell_{\nu} Sc^{-1/2},$
 where $\ell_{\nu}$ is the viscous dissipation cutoff.  
Assuming that the turbulent dynamics are governed by a direct enstrophy cascade (see also Sec.~\ref{sec:Reynolds}), 
the latter can be estimated as $\ell_\nu =\left(\nu^3/\langle \eta_\nu \rangle\right)^{1/6}$, where $\langle \eta_\nu \rangle$ 
is the mean enstrophy flux~\cite{boff}. 

All the dynamical equations, Eqs.~(\ref{ansa}-\ref{ansb},\ref{apza}-\ref{apzb}), are solved through the open-source code Basilisk 
(\url{http://basilisk.fr}).
The adopted boundary conditions are such that an inflow/outflow conditions are imposed 
on the left/right side of the domain, while free-slip conditions hold at the boundaries in the $y$-direction. 
On the obstacle we have a no-slip condition for the velocity while a no-flux condition is imposed for the two scalars, which are kept 
at the equilibrium values ($P_{eq},Z_{eq}$) at all sides of the domain.
Further details on the numerical approach and boundary conditions are provided in Appendix~\ref{app:num}.
For the initial conditions, we fix the longitudinal advecting velocity to the uniform inflow value $u_0$, while the transversal one is zero.
Following~\cite{hl}, the scalar fields are initially set to their equilibrium values. At a later time $t^*>0$, once the flow is in statistically stationary conditions, 
to mimic the arrival of a small phytoplankton population, we let a localized patch of $P$ density 
enter the system from the left of the island. Its spatial distribution is chosen to be of the form
\begin{equation}
P(\bm{x}, t^*) = P_{eq} + P_ae^{(- ((x-x_0)^2 + (y-y_0)^2)/w^2)}, 
\label{pini}
\end{equation}
where $P_a = 0.5$ is the amplitude of the excitation, $(x_0,y_0) = (-2,0.5)$ its location and $w=0.9$ its width, so $w \simeq l_0$.
The possible effect of the initial conditions will be soon explained.

\begin{figure}[ht]
	\includegraphics[width=\textwidth]{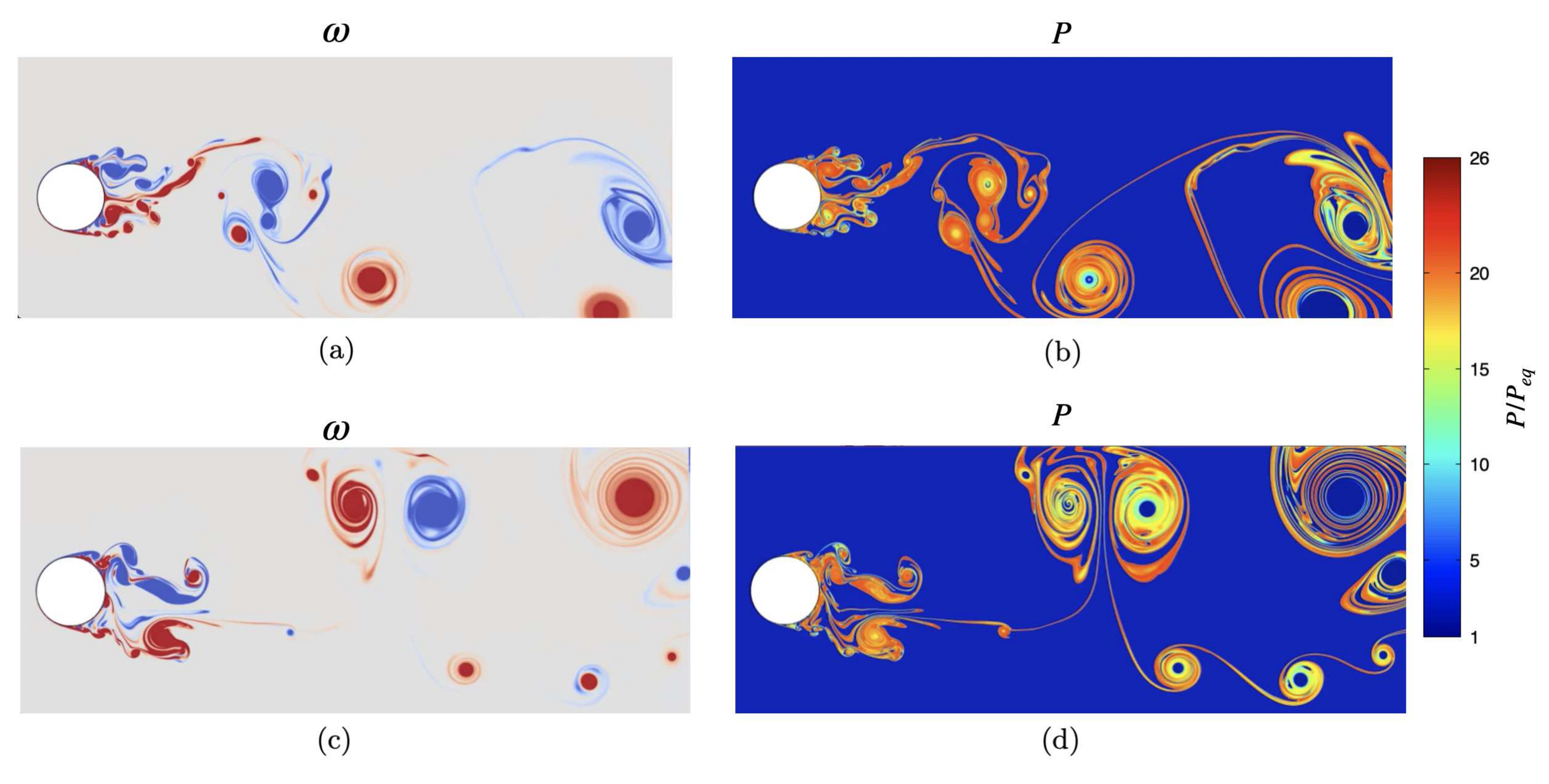}
	\caption{Vorticity field (a) and (c) and phytoplankton density (b) and (d) at $t/t_0=230$ (top panels) and $t/t_0=300$ (bottom panels) (with $t_0$ the dimensional value) at $Re = 20000$. In the left column the blue color stands for clockwise circulation  and the red color for
counterclockwise. The colorbar refers to the panels in the right column.}
	\label{fig:domain}
\end{figure}

In Fig.~\ref{fig:domain} we show some visualizations of the vorticity and phytoplankton fields for the simulation at the highest Reynolds number  considered. The planktonic patch is introduced at $t^*/t_0 = 110$  and it takes a certain time to overtake 
the obstacle and cover the domain with complex filamentary structures, as it has been pointed out in~\cite{neu}. 
The plankton patchiness follows the spatial organization of the flow: the correlation between the phytoplankton density and 
coherent structures is apparent in the figure and it will be quantitatively investigated in the following. 
It is here important to stress the transient character of the fluid motion: the flow structures, after spending some time in the vicinity of the island, continuously leave the domain through the right side.

\section{Results}
\label{sec:results}

\subsection{Flow regimes}
\label{sec:flow_regimes}

The flow past a cylindrical obstacle is a classical flow configuration in fluid mechanics. In its 2D version, it has already been considered 
as a relevant model to describe the wake behind an island to investigate the population dynamics of micro-organisms at the surface 
of the ocean, although through a prescribed streamfunction~\cite{Sandulescu_etal2007,Sandulescu_etal2008,BF2010,GF2020}. 
Let us recall that the flow becomes unsteady at moderate Reynolds numbers and for $40 < Re < 1000$ 
vortices are periodically shed from the obstacle, while for $Re>1000$, the separated flow becomes increasingly more turbulent displaying 
spatially and temporally irregular behaviour~\cite{vd}. 
In the periodic regime, a relevant non-dimensional parameter is the Strouhal number, $St = nd/U$, in which $U$ is the free-stream 
velocity intensity and $n$ the vortex-shedding frequency.  
The typical flow time scale is thus $T=1/n$, which gives the time interval between the appearance of two vortices of the same sign. 
For instance, at $Re=400$, in the absence of the biological scalar fields, we obtained a periodic flow, as the one used in kinematic simulations~\cite{Sandulescu_etal2007,Sandulescu_etal2008,BF2010,GF2020}, 
with $T \simeq 8$ in non-dimensional units, or $St \simeq 0.23$, 
in good agreement with experiments in a homogeneous non-rotating tank (where $St \approx 0.21$ \cite{z}). 
Increasing the $Re$ number, the flow becomes less and less regular and eventually the periodic behaviour is lost. 
In the present work, we have considered three simulations with the following Reynolds numbers: 
(A) $Re = 400$ ($N = 2^{11}$), (B) $Re = 2000$ ($N = 2^{12}$) and (C) $Re = 20000$ ($N = 2^{14}$), where 
we indicate in parentheses the maximum grid resolution $N$ for each case.

\subsection{PZ model}
\label{sec:biomodel}

The excitable character of the biological model can be appreciated by considering the temporal evolution of the system in the absence of flow 
($\bm{u}=0$ in Eqs.~(\ref{apza}-\ref{apzb})) with the described initial conditions for the patch (see Fig.\ref{fig:reac}). 
As in~\cite{tru}, here the outbreak is caused by a direct perturbation of $P$ via a sudden increase of its density. 
The response of the planktonic species, initially at their equilibrium density values $P_{eq}$ and $Z_{eq}$ everywhere in the domain, 
is characterized by a fast initial growth of the spatially averaged phytoplankton density, followed by a slower return to 
the equilibrium, caused by the (slower) growth of zooplankton. This picture is possible because the two species evolve on different time scales, 
given the presence of parameter $\gamma$ in Eq.~(\ref{Z_ode}) (or, analogously, in Eq.~(\ref{apzb})), which limits the zooplankton predation 
efficiency and thus allows phytoplankton to escape the $Z$ control to reach the carrying capacity. 
Therefore, we see that due to excitability plankton growth is a transient phenomenon in this system.


\subsection{Coupled biological and fluid dynamics}
\label{sec:fullmodel}

\begin{figure}[ht]
\captionsetup[subfigure]{labelformat=empty,justification=centering}
\captionsetup[figure]{justification=justified, singlelinecheck=off}
\begin{subfigure}{.5\textwidth}
\includegraphics[width=0.9\textwidth]{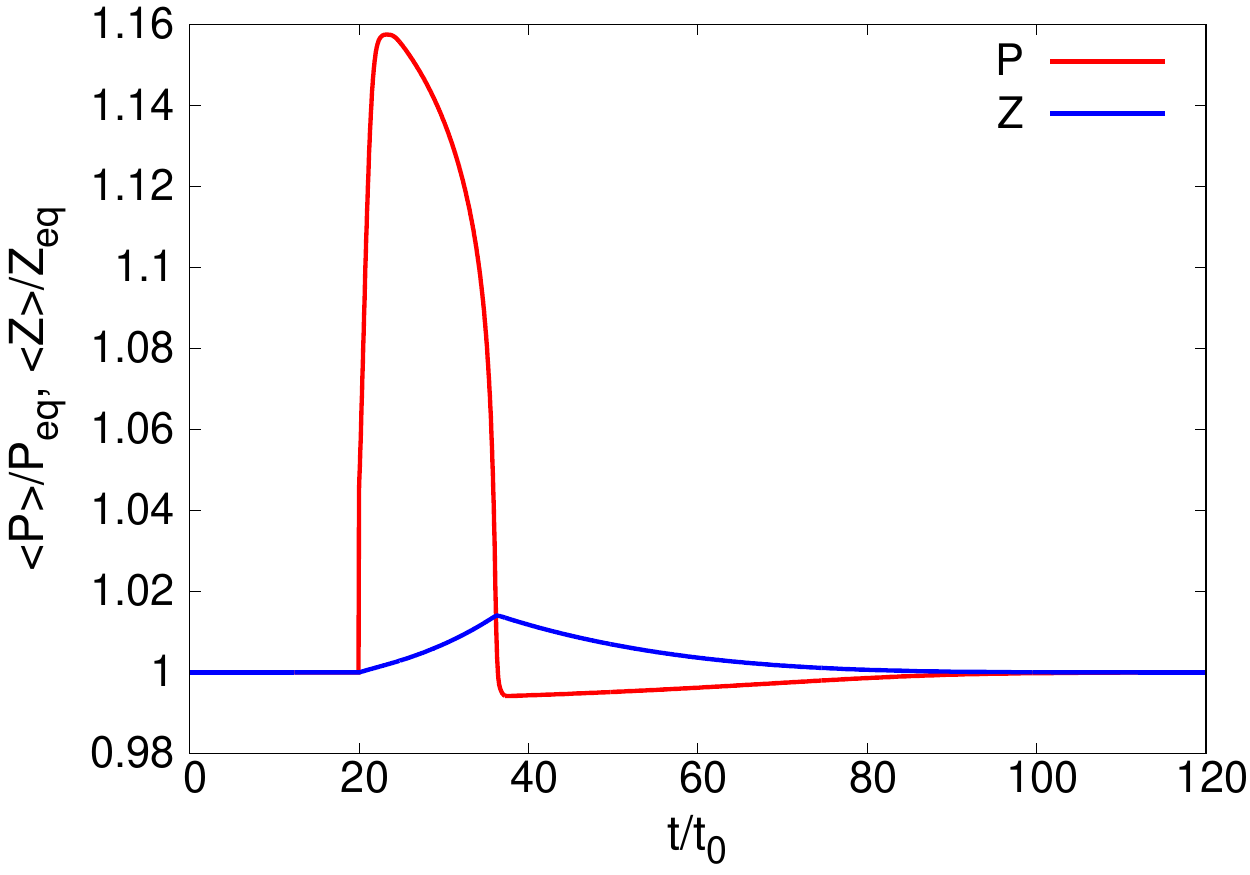}
\caption{(a)}
\label{fig:reac}
\end{subfigure}%
\begin{subfigure}{.5\textwidth}
\includegraphics[width=0.9\textwidth]{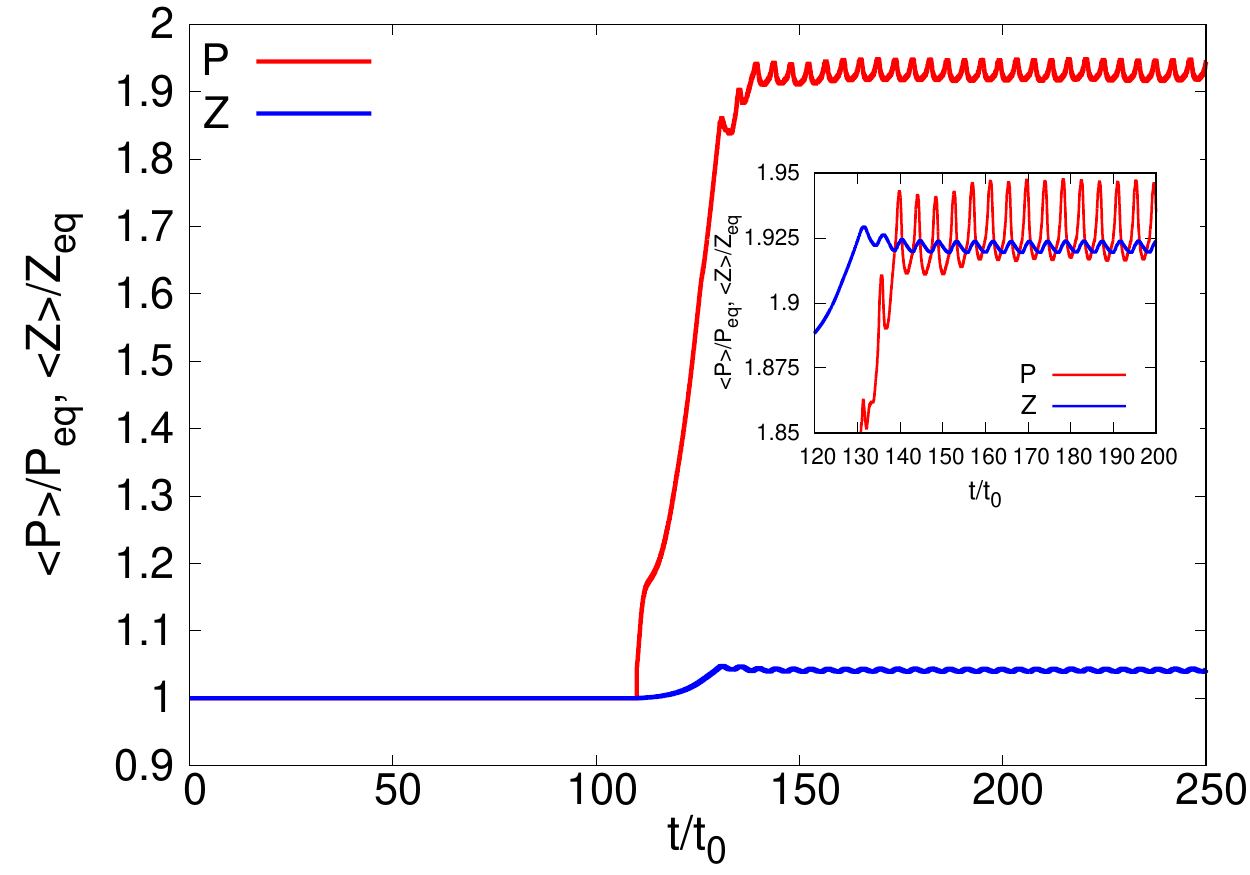}
\caption{(b)}
\label{fig:Rec}
\end{subfigure}%
\caption{Population densities of phytoplankton ($P$) and zooplankton ($Z$), 
 averaged in space and normalized by the corresponding equilibrium values, in the absence (a) and in the presence (b) of flow (at $Re=400$). 
In the inset of the right panel the $Z$ curve is shifted above, to superpose it to the $P$ curve, to highlight the delay of the 
zooplankton growth with respect to the phytoplankton one also in the presence of a flow.}
\end{figure}

Advection has remarkable consequences on the biological dynamics. 
As shown in Fig. \ref{fig:reac} at $Re=400$, the combined (transient) effects 
of fluid transport and population growth, give rise to a permanent excitation of the predator-prey system. 
Both $P$ and $Z$ now reach spatially averaged densities that are considerably larger than their equilibrium values. 
The temporal behaviour is dictated by the vortex shedding, with period $T\simeq 8$ (in non-dimensional units), 
to which the biological dynamics is slaved. 
Indeed, as shown in Fig.~\ref{fig:Rec}, both populations oscillate regularly in time with period $T$. 
This result confirms the outcomes of previous works~\cite{hl,lo,neu}, where the same change of behaviour was investigated using kinematic flows 
(in slightly different configurations) and a basic mechanism was proposed in terms of the existence of a chaotic saddle, 
namely a flow structure generated by the presence of the obstacle, in which fluid parcels remain trapped for very long time.

As suggested by dynamic visualizations (see discussion in Appendix~\ref{app:video}), the plankton growth appears here to start in the boundary layer around the obstacle, rather than just in the chaotic saddle after it. 
Indeed, the phytoplankton perturbation is introduced in front of the obstacle and encounters the boundary layer, where the residence-time is long.
That might indicate different possible mechanisms triggering the bloom, or a dependence on initial conditions. 
To address this issue, several tests have been performed to study the impact of the specific form of the $P$
initial condition. 
The system response turned out to be independent of the amplitude $P_a$ of the initial perturbation (no appreciable differences arise by varying this parameter in the range 
$0.1 \leq P_a \leq 0.5$), as well as on its size $w$. 
Regarding its location, \ie $(x_0,y_0)$, the algal bloom occurrence and its permanent character have been found 
for different values of $x_0$: initializing the perturbation in front of the obstacle or behind it does not affect the 
dynamics of the scalar fields, provided that $\vert{x_0} \vert< 2.5d$. For initial positions further 
downstream of the obstacle, the patch is advected away by the flow without giving rise to a permanent excitation.
Our results show therefore that the characteristics of the bloom are largely independent 
of the precise mechanism trapping plankton, provided the hydrodynamic time-scale is much larger than the biological one.
This intriguing result motivated us to study also various laminar cases at $2\lesssim Re \lesssim 10$ (results not shown for the sake of brevity), since these flows are not chaotic but they have two stagnation points in front of and behind the obstacle.
While the results are globally different from the turbulent ones, as there is no mixing, the plankton still displays the same quantitative growth as in the chaotic case.

\begin{figure}[ht]

	\includegraphics[width=0.45\textwidth]{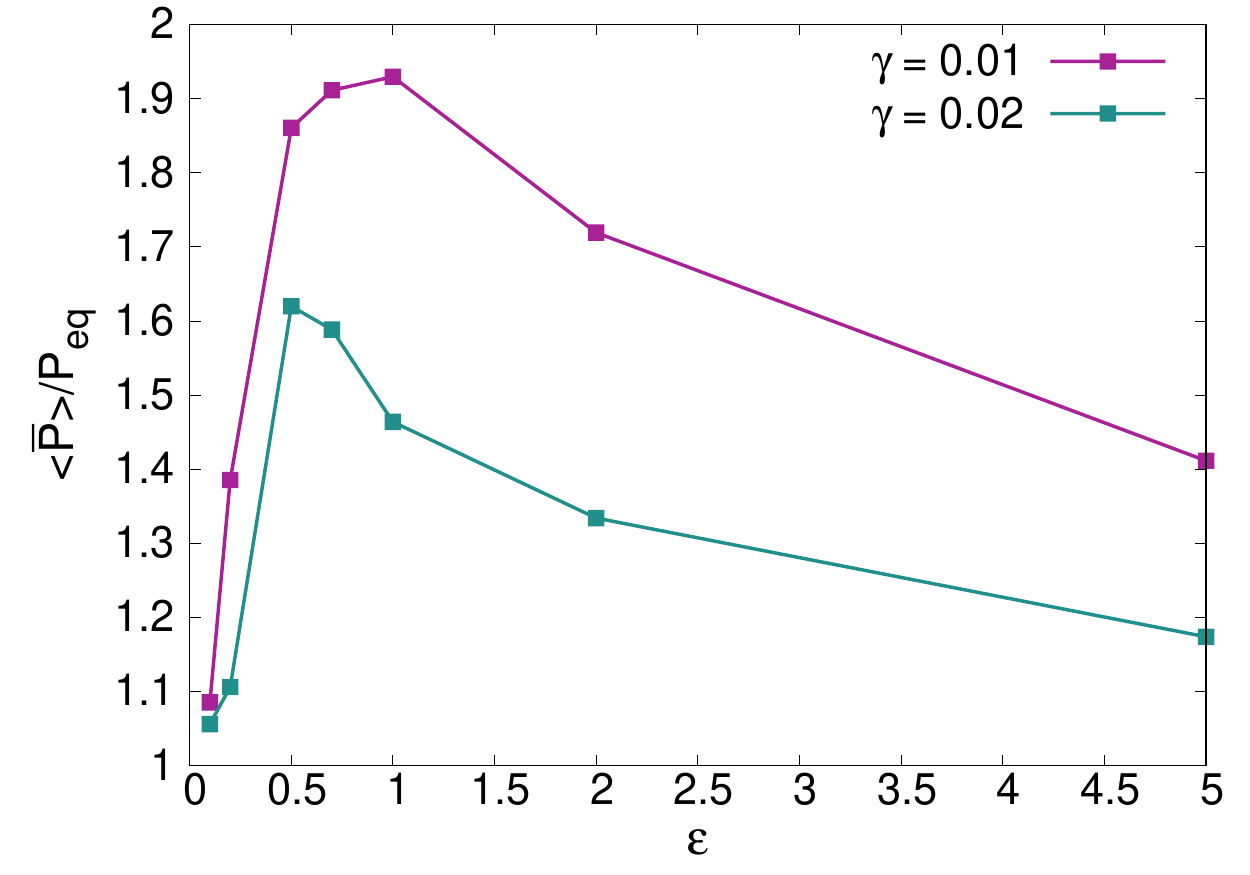}
	\caption{Phytoplankton population, averaged in space and time, as a function of $\epsilon$, 
        for two values of $\gamma$ at $Re=400$.}
	\label{fig:epsilon}
\end{figure}

Then, we have investigated through several simulations the impact
of the biological properties, namely the ratio of the transport to the biological activity time scales $\epsilon$, and the ratio of biomass consumed to biomass of new herbivores produced, \ie the predation efficiency $\gamma$. 
The results are reported in Fig.~\ref{fig:epsilon}.  
The dependence of the spatially and temporally averaged $P$ on $\gamma$ is evident: no matter the value of $\epsilon$, 
the mean global value of phytoplankton $\langle \overline{P} \rangle$ is higher for $\gamma = 0.01$ than for $\gamma=0.02$, 
\ie for a smaller predation efficiency. 
We checked that such a feature is general, considering different increasing values of $\gamma$. 
Beyond a limiting value $\gamma = 0.05$, the grazing by the zooplankton dominates and the system dynamics cannot 
sustain a permanent excitation. The phytoplankton bloom is in that case only a transient event, as in the absence of flow. 
The impact of $\epsilon$ is more complex, as also suggested by~\cite{hl}. 
Independently of $\gamma$, an optimal value $\epsilon^*=O(1)$ of this ratio exists. 
It is attained when the transport by the flow is slower than phytoplankton growth (as $\beta>1$, \ie $t_0>r^{-1}$, and $\epsilon \simeq 1$). 
For values of $\epsilon$ much smaller or much larger than $\epsilon^*$, we observe a tendency 
to recover the equilibrium state, which can be understood as follows.  
Decreasing $\epsilon$ corresponds to making advection faster (or growth slower). 
In this case, the initial phytoplankton patch is deformed by stirring and diffusion but soon decays 
downstream of the obstacle, as the biological dynamics are too slow to sustain its growth. 
In the opposite limit of large $\epsilon$, corresponding to very slow flow (or very fast growth), 
a sudden excitation occurs upstream of the island but the two populations start to get back to their equilibrium states 
before reaching the obstacle and so when they are entrained in the wake the initial abundance of phytoplankton has been already partially consumed, resulting in a smaller value of the average biomass in the system.
The behaviour for $\epsilon \ll \epsilon^*$ is consistent with the transition to permanent excitation previously detected 
using a kinematic perturbed jet~\cite{hl}. The behaviour at $\epsilon \gg \epsilon^*$, instead, points to 
a second transition, to de-excitation, which could not be found in~\cite{hl}, likely due to the specific topology of the flow employed, 
but which was documented for a different excitable medium in the presence of 
a blinking vortex-sink flow~\cite{NLHP2002}.

\subsection{Role of turbulence and impact of Re}
\label{sec:Reynolds}

\begin{figure}[ht]
\captionsetup[subfigure]{labelformat=empty,justification=centering}
\begin{subfigure}{.5\textwidth}
\includegraphics[width=0.9\textwidth]{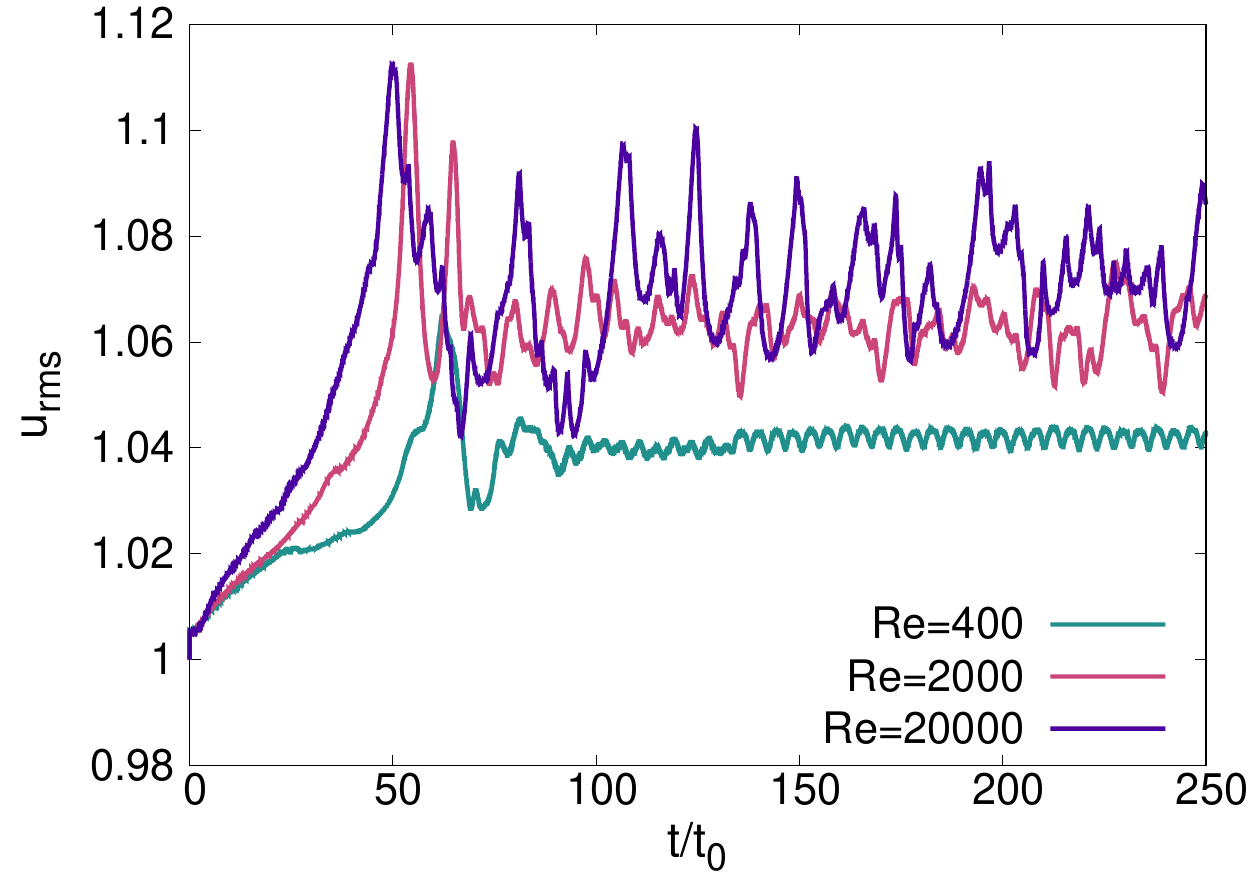}
\caption{(a)}
\label{fig:Rev}
\end{subfigure}%
\begin{subfigure}{.5\textwidth}
\includegraphics[width=0.9\textwidth]{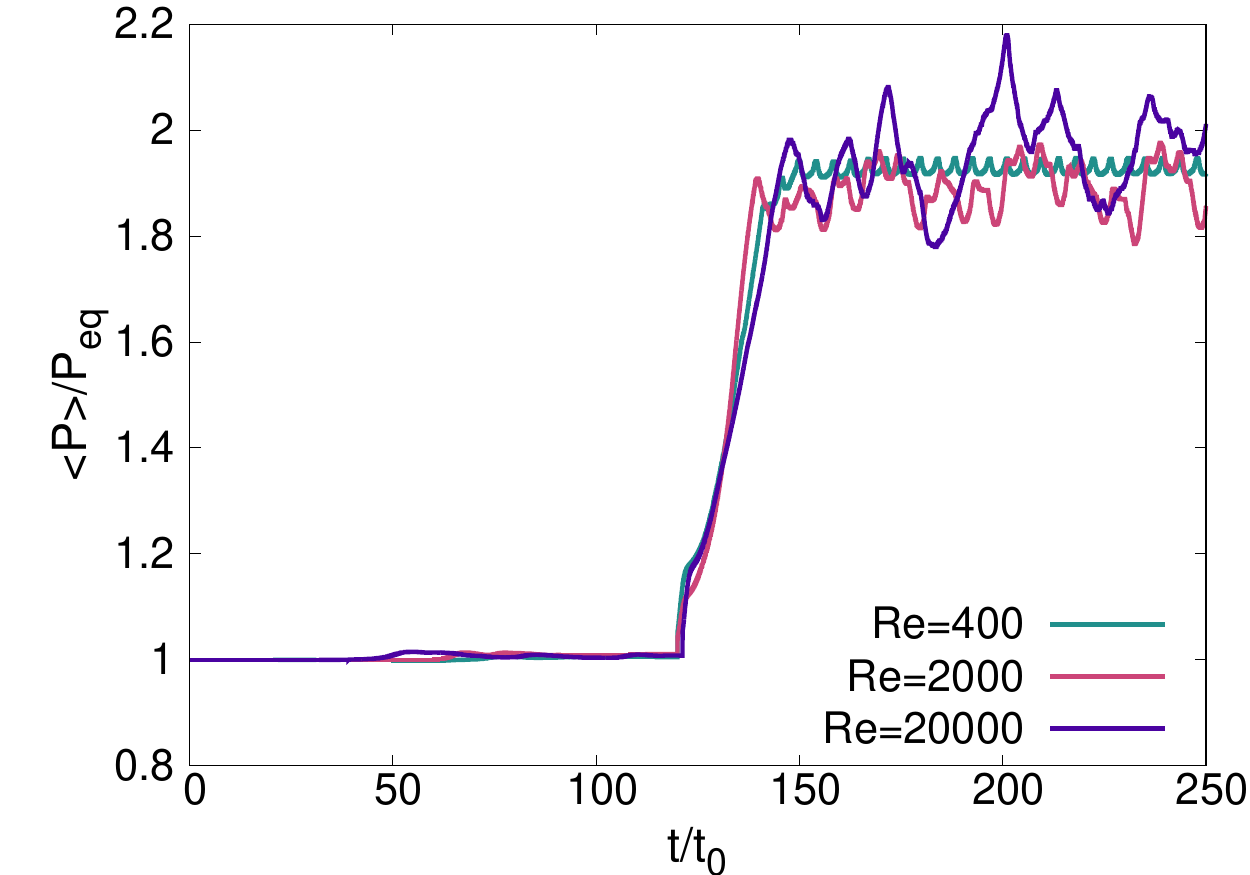}
\caption{(b)}
\label{fig:c1b}
\end{subfigure}%
\caption{Root mean square fluid velocity (a) and spatially averaged phytoplankton density  (b) versus time for increasing $Re$ numbers.}
\label{fig:vrms}
\end{figure}

In a realistic geophysical configuration, the Reynolds number is typically huge ($10^8 \lesssim Re \lesssim 10^{10}$). 
While it is not possible to carry out fully resolved simulations of flows at such large values of $Re$, 
it is crucial to understand if the basic features detected at moderate $Re$ are robust with respect to the increase of this control parameter.
We investigated the role of the turbulent dynamics using the fixed values $\epsilon=1$ and $\gamma=0.01$ of the coupling parameters discussed 
in Sec.~\ref{sec:fullmodel}. As it can be seen in Fig.~\ref{fig:vrms}, when $Re \geq 2000$ the root mean square (rms) flow intensity 
$u_{\mathrm{rms}}$ looses its periodicity and displays a highly irregular temporal behaviour. As already observed from Fig.~\ref{fig:domain}, at $Re=20000$ the vortices 
do not travel along a straight path but are now deflected in the transversal direction, above and below the center line. Due to the spatial and temporal complexity of the flow, the planktonic populations no longer oscillate periodically in time and the filamentary structures appear thinner and more convoluted, especially close to the island. 
This feature is reflected in the measure of the globally averaged population density, $\langle P\rangle$ (~\ref{fig:c1b}), 
for which the amplitude and irregularity of fluctuations are modified by the increased chaoticity of the flow, in spite 
of a weak dependence of its (temporal) mean value on $Re$.

In the following, we present the results of a deeper analysis about the local properties of plankton fields. 
In particular, we aim at investigating their correlation with the underlying flow. 
We focus on simulation $C$ (at $Re=20000$), representative of the fully turbulent regime, but we will also perform some comparisons  
with the results in the periodic regime of simulation $A$ (at $Re=400$).  

\begin{figure}[ht]
\captionsetup[subfigure]{labelformat=empty,justification=centering}
\begin{subfigure}{.5\textwidth}
\includegraphics[width=0.8\textwidth]{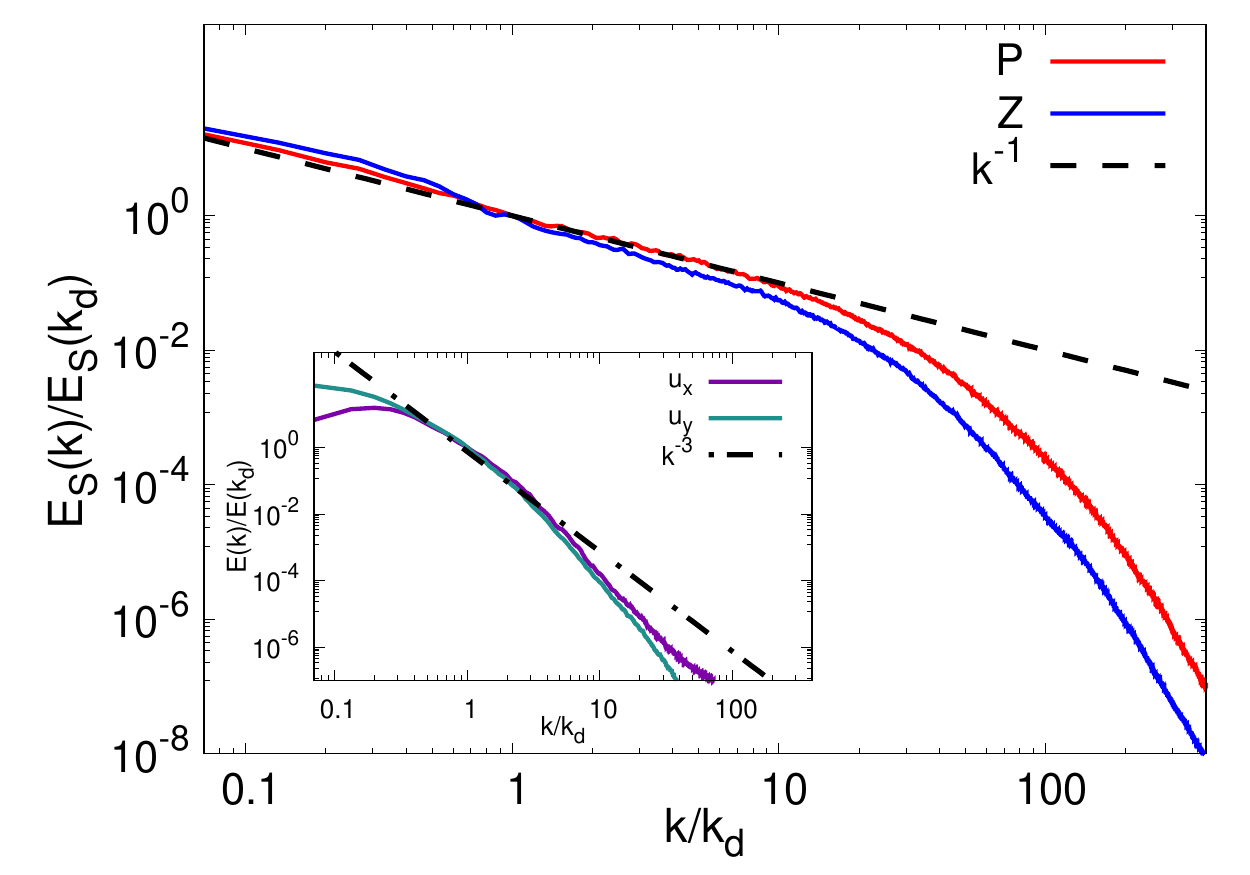}
\caption{(a)}
\label{fig:spea}
\end{subfigure}%
\hfill
\begin{subfigure}{.5\textwidth}
\includegraphics[width=0.8\textwidth]{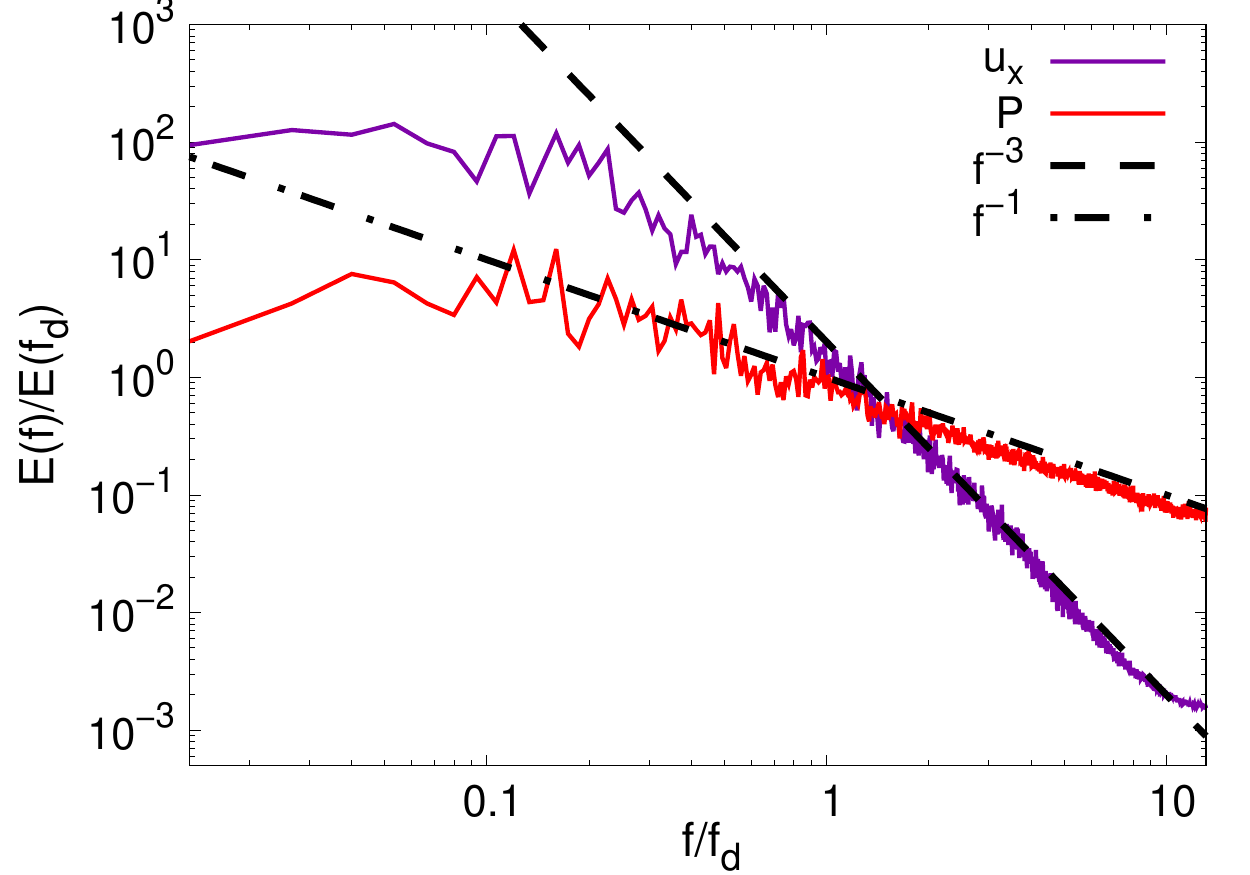}	
\caption{(b)}
\label{fig:speb}
\end{subfigure}%
\caption{ (a) Spatial spectra of 
phytoplankton and zooplankton fluctuations, and spectra of velocity components fluctuations (inset) at $Re=20000$. 
All these spectra are normalized by the value corresponding to $k_d$, the wavenumber 
associated with the island diameter $d$.
The spectra are computed along the $y$-direction and then averaged for $1.5d \leq x \leq 10d$ and $150 \leq t \leq 300$. 
(b) Power spectra, in the frequency domain, of longitudinal velocity $u_x$ and phytoplankton fluctuations at $Re=20000$, 
normalized by the value corresponding to $f_d$. Note that $f_d=n/St$, where $n$ is the vortex-shedding frequency.}
\label{fig:spectra}
\end{figure}

We first consider the spectra of  scalar variance and kinetic energy.
Figure~\ref{fig:spea} reports one-dimensional spectra of velocity ($u_x,u_y$) and scalar ($P,Z$) fluctuations,   
in the transverse direction (with respect to the mean flow), 
computed in the subdomain $1.5d \leq x \leq 10d$ (with $d$ the island diameter).  
Note that the Fourier transform is performed along the $y$ direction at several fixed longitudinal positions $x$ and that the resulting spectra are subsequently  
averaged in the $x$ direction. 
To improve the statistics, we further perform a temporal average of the latter, in the temporal interval $150 \leq t \leq 300$ (in non-dimensional units), where the system has reached a statistically stationary state. 
Energy spectra are compatible with a scaling $E(k) \sim k^{-3}$ over approximately one decade, pointing to the existence of a smooth flow and 
a direct enstrophy cascade~\cite{kra}, with an injection scale $k_d$ corresponding to the diameter of the obstacle. 
At the highest wavenumbers they tend to be steeper, when dissipation starts dominating.
We remark that, for our discussions about the interplay between fluid and biological dynamics, the precise slope 
of $E(k)$ is not expected to play a major role, as long as the spectrum is steep enough. 
In fact, both for a spectrum scaling as $k^{-3}$ and for a steeper one, the flow possesses a single time scale, 
determined by the strain (which essentially acts at the largest scales, in both these cases).

The wavenumber spectra of plankton populations, and particularly that of phytoplankton, are commonly used 
to characterize biological patchiness. Several experimental measurements, obtained using transects 
from research vessels and satellite images of the sea color (a proxy for phytoplankton concentration), have been performed 
in different regions~\cite{DP1976,Smith_etal_1988,MS2002,LK2004,franks}. 
While they all provide evidence of 
a power-law behaviour $\sim k^{-n}$, different values of the exponent are reported, ranging from $n=1$ to $n=3$, 
which suggests that different physical and biological processes may be relevant for plankton spatial variability, 
in different conditions. In our simulations, for
both planktonic species we find a spectrum close to $ E_S(k)\sim k^{-1}$ in a wide range extending for approximately 
two decades, starting from the largest scale. At the smallest scales, dissipative effects appear to dominate.
We have therefore found that the energy distribution of passive reactive scalars is the same as that encountered 
for passive non-reactive ones \cite{batchelor1959small} (more details in Appendix~\ref{app:comp}).
The observed behaviour is in agreement with theoretical predictions for two interacting species evolving in a 2D smooth turbulent flow~\cite{powell}. 
This result implies that population dynamics somehow reduce patchiness at large scales, giving rise to ``whiter'' spectra of the scalar fluctuations, meaning flatter than for the 
turbulent flow. 
In Figure~\ref{fig:speb}, we also show the power spectra in the frequency domain for phytoplankton and the longitudinal velocity component, 
which confirm the above scaling behaviour even more accurately. The Fourier transform is here performed in the time domain by collecting the data at a distance 
$x=5d$ downstream of the obstacle and then averaging the resulting spectra in the $y$ direction.

\begin{figure}[ht]
	\includegraphics[width=0.475\textwidth]{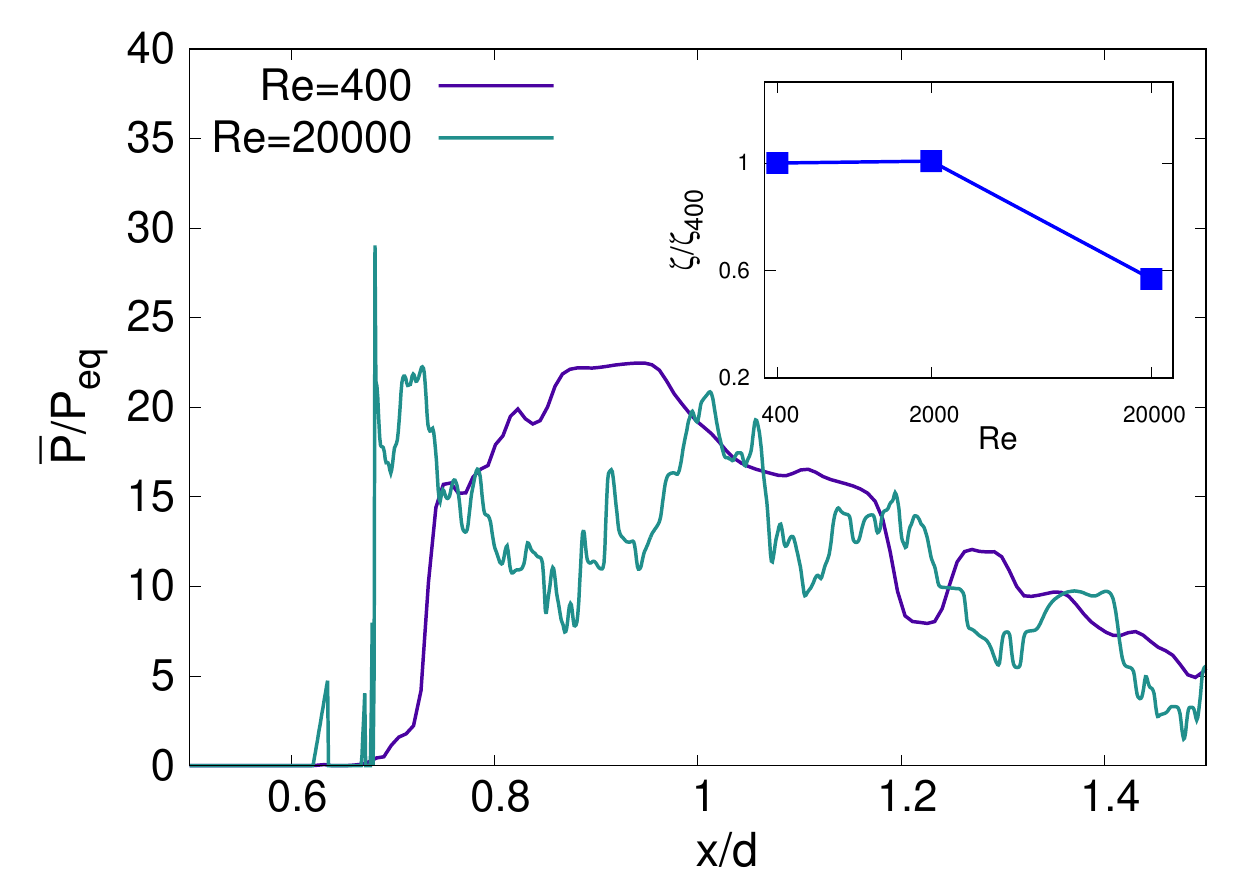}	
	\caption{Transects of time-averaged phytoplankton density, normalized by the equilibrium value,
        at the center line $y/d=0$ for $Re=400$ and $Re=20000$. 
        Inset: ratio $\zeta=\beta_{eff}/\langle \eta_\nu \rangle^{1/3}$ (see text), normalized by its value at 
       $Re=400$ for the three simulations ($Re=400,2000,20000$).}
      	\label{fig:comp}
\end{figure}

Having quantified the scale-by-scale energetic content of the planktonic populations, we now turn to their spatial features.  
Indeed, the characteristic spatial patterns of the $P$ and $Z$ fields are a consequence of the time scales over which they respond 
to changes in their environment caused by turbulent advection~\cite{abra}. 
While global quantities of the planktonic species share a similar qualitative behaviour
in all our simulations, see Fig.~\ref{fig:vrms}, the amplitude and frequency of the fluctuations grow with $Re$. 
To inspect the effect on spatial features of increasing the Reynolds number, we can consider a horizontal cut along the line $y/d=0$ to obtain a transect of the phytoplankton density in the vicinity of the obstacle. 
As it is evident from Fig.~\ref{fig:comp},  while for $Re=400$ the population density field appears quite smooth, for $Re=20000$ it appears considerably more jagged, displaying much stronger gradients. 
Some insight on this change of behaviour can be gained by adapting a criterion, originally 
developed in the framework of linear-decay chemical reactions (with rate $b$) in laminar flows, to identify the so-called 
smooth-filamental transition~\cite{neufeld}.  In a nutshell, the theory, constructed in a Lagrangian reference frame, is based on the comparison 
between the largest Lyapunov exponent of the flow $\lambda$ and the chemical (biological) Lyapunov exponent ($b=\lambda_c$), 
quantifying the exponential growth rate of the chemical concentration (biological population density), 
which allows to obtain the H\"older exponent $\zeta$ of the reactive field as:    
\begin{equation}
\zeta = \mathrm{min} \left\{\frac{b}{\lambda}, 1 \right\}.
\end{equation}
If $\zeta = 1$ the field is smooth (differentiable), while for $0 < \zeta < 1$ it is filamental.

In the context of the present study, a simple possibility to estimate the corresponding quantities is by dimensional arguments. 
The Lyapunov exponent of the flow should be proportional to the inverse of the fastest fluid time scale~\cite{ruelle}.
Consistent with the occurrence of a direct enstrophy cascade (see Fig.~\ref{fig:spectra}), 
we then take $\lambda \sim \langle \eta_\nu \rangle^{1/3}$ (with $\eta_\nu$ the enstrophy flux). 
For the biological dynamics we consider an effective phytoplankton growth rate 
$\beta_{eff}=\partial_t \langle P\rangle/\langle P\rangle$, computed in the early growth regime ($110 \leq t \leq 130$ 
in non-dimensional units) before the statistically steady state.
In the inset of Fig.\ref{fig:comp} our estimation of $\zeta$, normalized by the value at the smallest 
Reynolds number $\zeta_{400}$, is plotted versus $Re$ for simulations $A$, $B$ and $C$. 
In spite of the limitations of our dimensional approach, the criterion turns out to be  effective in capturing the transition.  
A clear decrease is found between the first two values of $Re$, for which the ratio is close to unity, and the largest one, 
for which it is definitely smaller.
We then conclude that the filamentary distribution of phytoplankton is a direct consequence of the increasing turbulent character of the flow.

\begin{figure}[htb]
\captionsetup[subfigure]{labelformat=empty,justification=centering}
\begin{subfigure}{.5\textwidth}
\centering
\includegraphics[width=1.\textwidth]{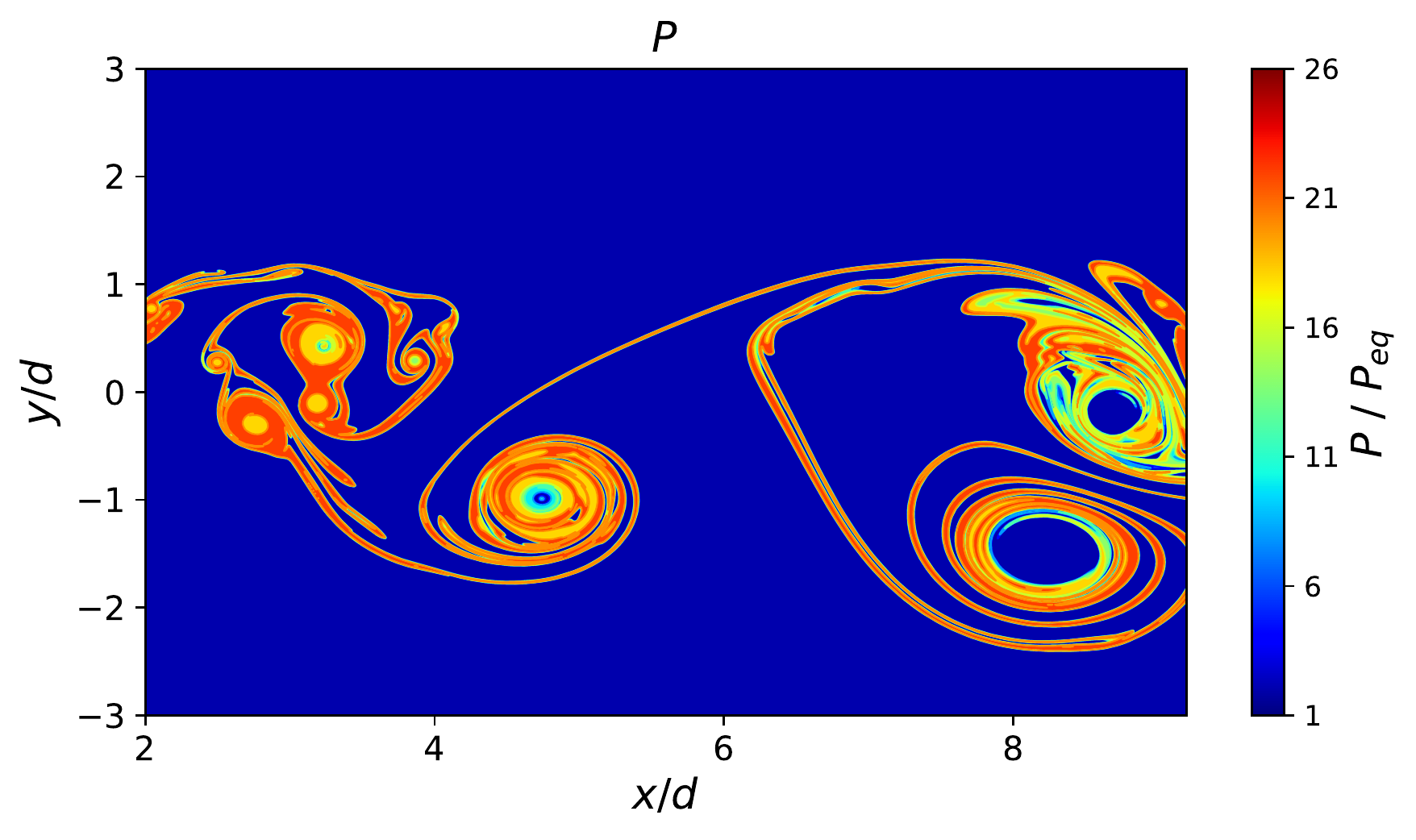}
\caption{(a)}
\label{fig:p}
\end{subfigure}%
\begin{subfigure}{.5\textwidth}
\centering
\includegraphics[width=1.\textwidth]{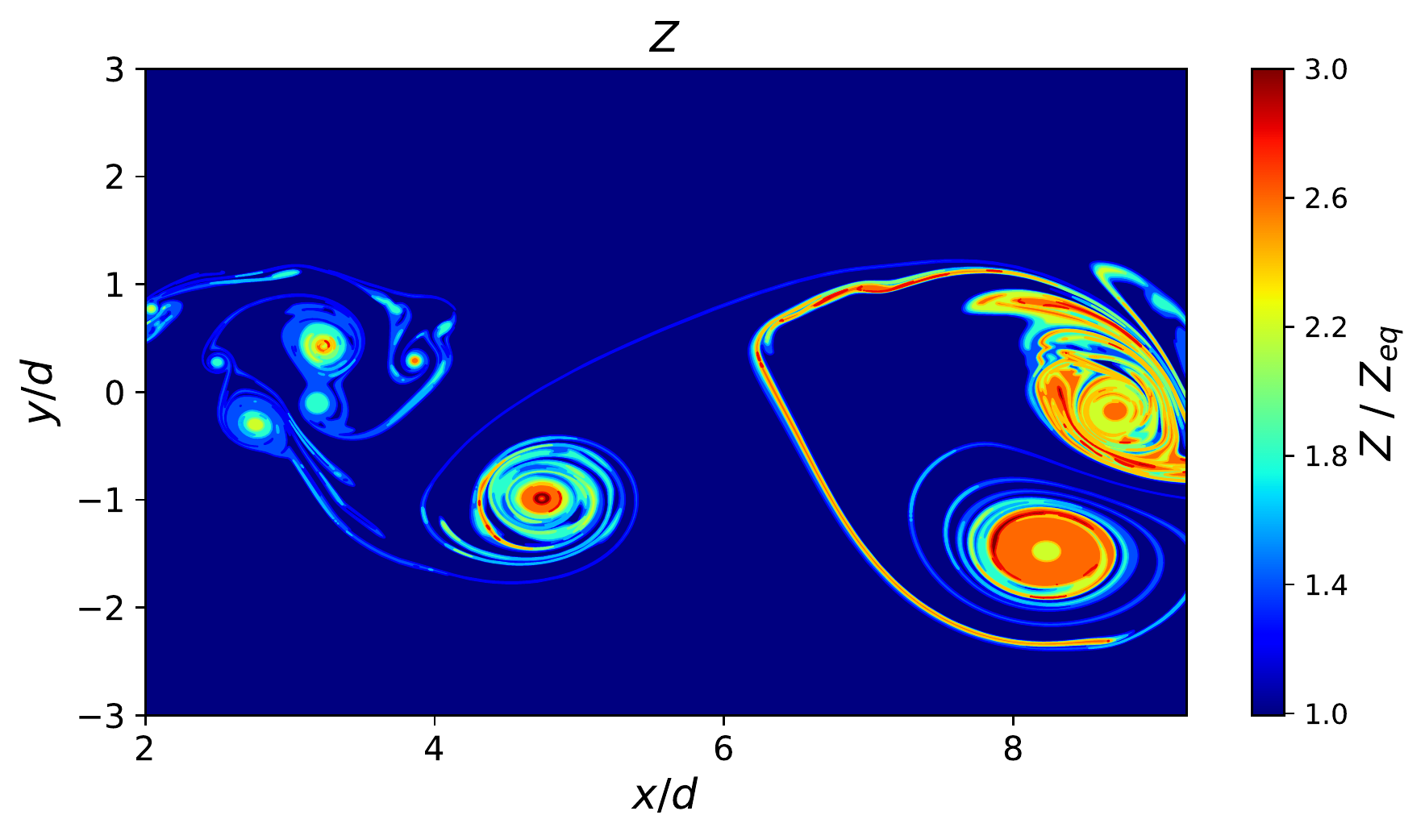}
\caption{(b)}
\label{fig:z}
\end{subfigure}\par\bigskip
\centering
\caption{Instantaneous fields of phytoplankton density  $P$ (a) and zooplankton density $Z$(b),
for $Re=20000$ at $t = 230$ (in non-dimensional units), in the statistically steady state.}

\label{fig:fields}
\end{figure}

Based on the previous observations about the relationship between the statistical features of the turbulent and biological dynamics, 
it is now interesting to explore how the flow and scalar fields are correlated in space, in terms of structures.
For this purpose, at a given time in the statistically stationary state, we analyze both the phytoplankton 
and zooplankton spatial distribution at $Re = 20000$ 
in Fig.~\ref{fig:fields}.
As it has been already observed in Fig.~\ref{fig:domain}, both $P$ and $Z$ fields display structures similar to those present in the vorticity field.  
Very close to the island, several vortices of different sizes are present, somehow connected by filaments. 
Further downstream of the obstacle, such vortices grow and interact among each other, giving rise to two large and more 
separated eddies.
Phytoplankton winds around vortices, mainly concentrating in filaments, immediately downstream of the obstacle, 
and it progressively leaves the vortex cores, in which $P$ is at the equilibrium value $P_{eq}$ 
(Fig.~\ref{fig:p}). The zooplankton distribution parallels that of $P$, in terms of correlation between its extremal values 
and flow structures (Fig.~\ref{fig:z}). However, maxima of $Z$ correspond to minima of $P$, as \eg in eddy cores. 
The resulting picture is therefore due to two combining effects. The flow structures entrain the two scalars and transport 
them across the domain. The predator-prey biological interactions determine, locally, the relative abundance of the two species: 
where zooplankton grows, it consumes phytoplankton; where $Z$ is absent, instead, $P$ can grow more.

Similar results were previously found coupling the NPZ model (including nutrient dynamics) with a kinematic flow model~\cite{Sandulescu_etal2008}, essentially in the same geometrical configuration. 
Still for inflow densities at the equilibrium values, it was reported that phytoplankton tends to be localized mostly 
in the periphery of vortices, as in our study, which seems indicative of the generality of fluid transport 
and stirring mechanisms leading to filament formation. 

A more quantitative analysis of the correlation between the flow and the phytoplankton distribution 
was performed by considering an Eulerian quantity, the turbulent kinetic energy ($\mathcal{K}_E$), 
which is strictly linked to the flow stirring, since it is expected that more energetic turbulent areas would also present stronger horizontal stirring~\cite{rossi2008comparative,rossi2009surface}.
Denoting $u_i'(\bm{x},t) = u_i(\bm{x},t) - \overline{u}_i(\bm{x})$ 
(with $i=x,y$) the components of the fluctuating velocity field and with the overbar the temporal average over the time interval $t \in [150,300]$, 
the $\mathcal{K}_E$ field is given by:
\begin{equation}
\mathcal{K}_E(\bm{x},t) = \frac{1}{2} {\left( u_x'(\bm{x},t)^2 +  u_y'(\bm{x},t)^2 \right)} \, .
\label{eq:tke}
\end{equation}

\begin{figure}[htb]
\centering
\includegraphics[width=0.5\textwidth]{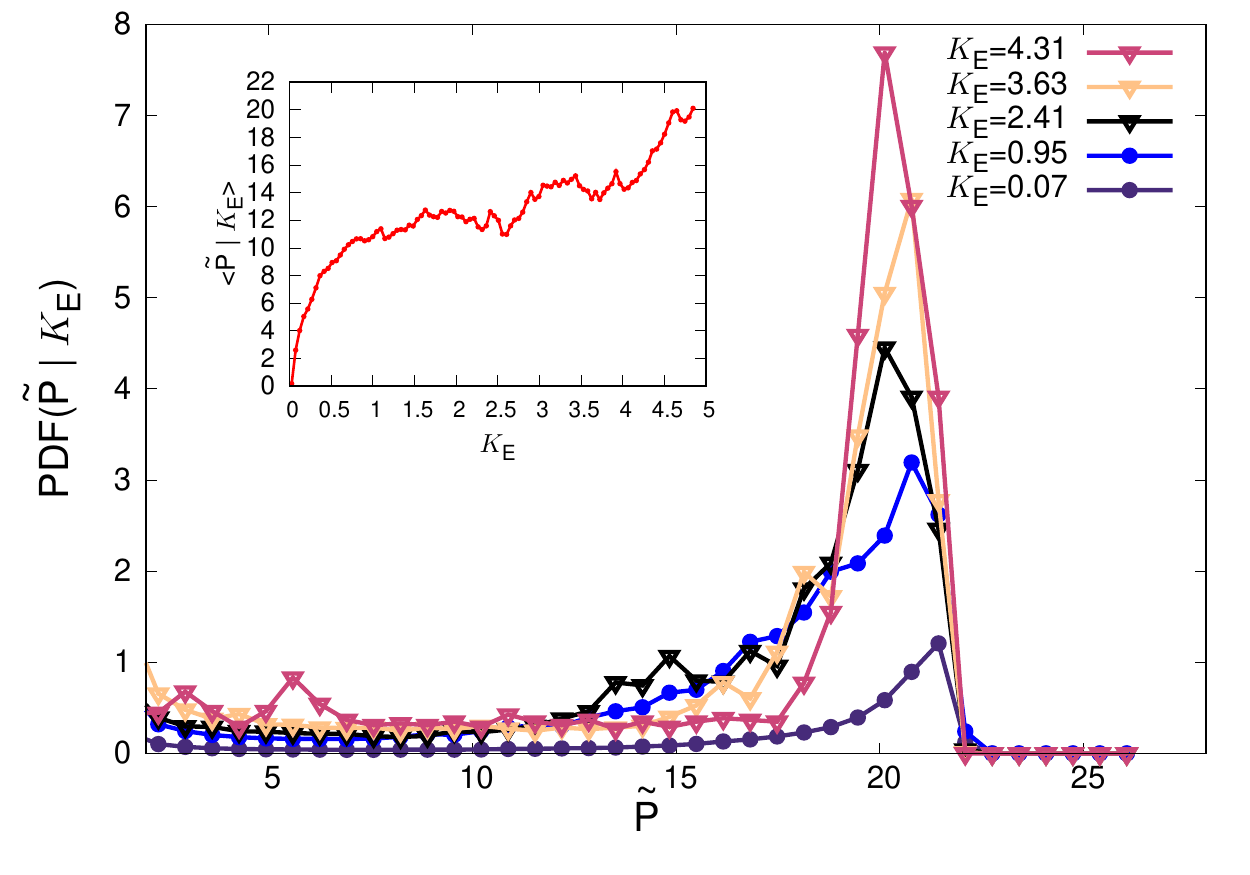}
\caption{PDF of $\tilde{P} = P/P_{eq}$ conditioned on $\mathcal{K}_E$ at $Re = 20000$. The different curves correspond to the PDF conditioned on increasing values of $\mathcal{K}_E$, as reported in the legend. 
In the inset the conditioned average of $\tilde{P}$ is shown as a function of the $\mathcal{K}_E$.}
\label{fig:pdf}
\end{figure}

Similarly to what is done in~\cite{hernandez2012seasonal}, we compute the average and the probability distribution function (PDF) 
of $\tilde{P} = P/P_{eq}$ conditioned on $\mathcal{K}_E$, over the time interval $t \in [150,300]$.

As shown in Fig.~\ref{fig:pdf}, we find that the phytoplankton distribution and 
the turbulent kinetic energy are positively correlated: 
the PDF conditioned on $\mathcal{K}_E$ exhibits a peak, for a value of $P$ essentially independent of $\mathcal{K}_E$, whose height 
increases with $\mathcal{K}_E$. The conditioned average shows a growth trend as a function of $\mathcal{K}_E$, which means that phytoplankton tends to concentrate in regions associated to high values of $\mathcal{K}_E$, i.e. characterized by intense flow stirring. \\
The above evidences then suggest that the stirring intensity plays a primary role in determining the plankton distribution in space and, indeed, in the present configuration it promotes biological productivity.


\subsection{Effect of the obstacle shape}
\label{sec:shape}

Real islands clearly do not have a perfectly circular shape and their boundaries are not necessarily smooth. 
Indeed, rocky shorelines are known to be well described by fractal curves~\cite{Mandelbrot1967,BCDS2008}. 
Therefore, from a general perspective, it is interesting to study the impact on the dynamics of a rough surface delimiting 
the island. To this aim, we repeated the simulation at $Re=400$ by replacing the circular obstacle 
with a geometrically irregular shape characterized by a rough boundary, expressed in terms of a truncated Steinhaus series~\cite{rough}:
\begin{equation}
\rho(\theta) = \rho_0 + A\sum_{k=1}^{\mathcal{N}}(-1-p)^{1/2}k^{p/2}\cos{(k \theta + \phi_k)},
\label{eq:rough}
\end{equation}
with $\theta \in [0,2\pi]$, $\rho_0 = l_0$ (the average radius), $\mathcal{N} = 1000$, $p=-2$ and $\phi_k$ independent random variables uniformly 
distributed in $[0,2\pi]$.
The function in Eq.~(\ref{eq:rough}) has a power spectral density for all non-zero wavenumbers $k$ that decays as $\sim k^p$, 
where $p=-2\xi-1$, and is fractal in the limit $\mathcal{N} \to \infty $, with fractal dimension $D_f = 2 - \xi$.
In the present case, to have an effect on the turbulent dynamics, the value of $A$ should be chosen such that 
the corrugations described by $\rho(\theta)$ cross the boundary layer around the obstacle. 
As the width of the latter can be estimated as $\sqrt{\nu l_0/u_0} \approx 0.1$ (in non-dimensional units), 
we chose two increasing values of $A$, \ie $A=0.15$ and $A=0.6$.
\begin{figure}[htb]
\captionsetup[subfigure]{labelformat=empty,justification=centering}
\begin{subfigure}{.5\textwidth}
\centering
\includegraphics[width=0.9\textwidth]{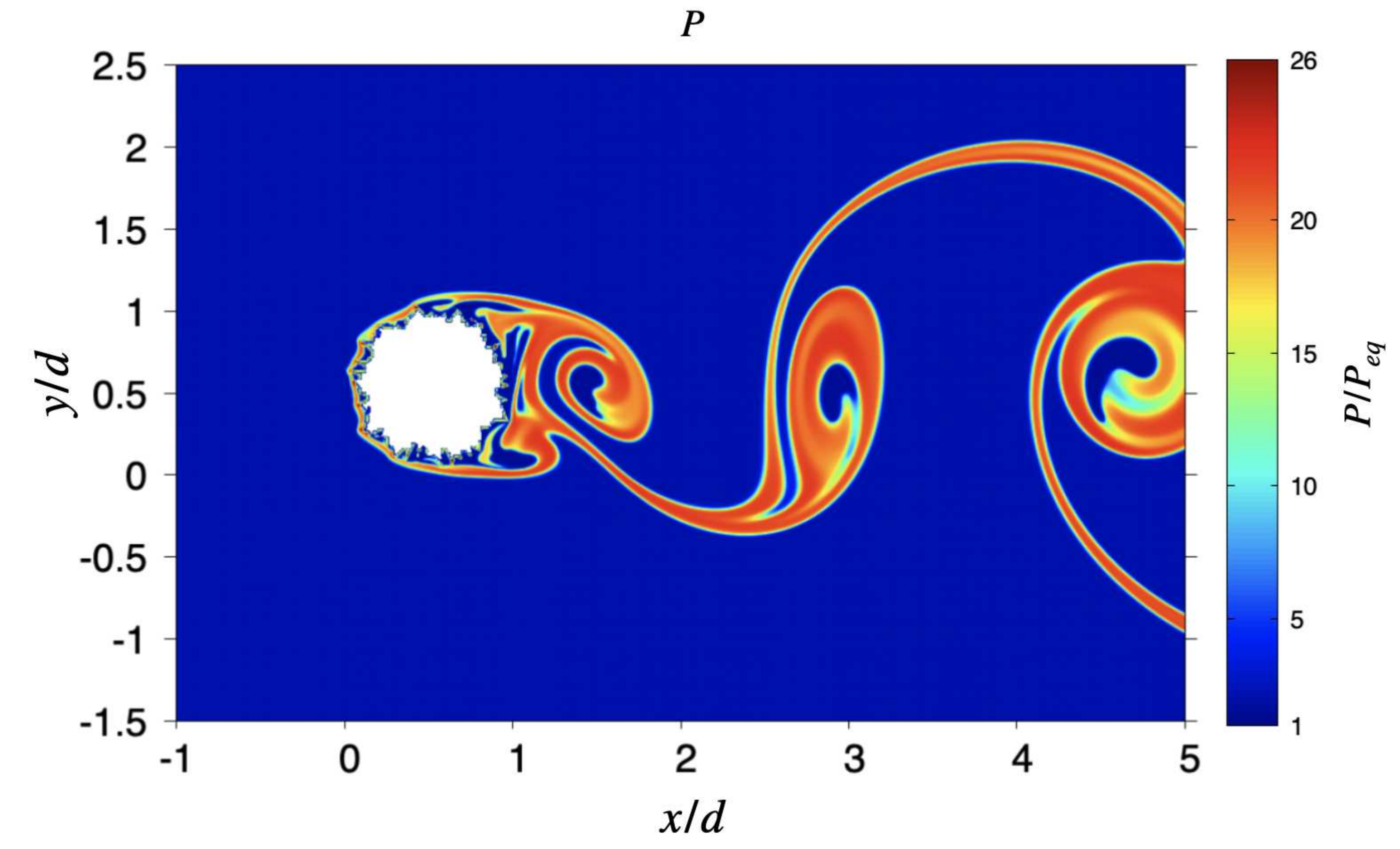}
\caption{(a)}
\label{fig:shapea}
\end{subfigure}%
\begin{subfigure}{.5\textwidth}
\centering
\includegraphics[width=0.8\textwidth]{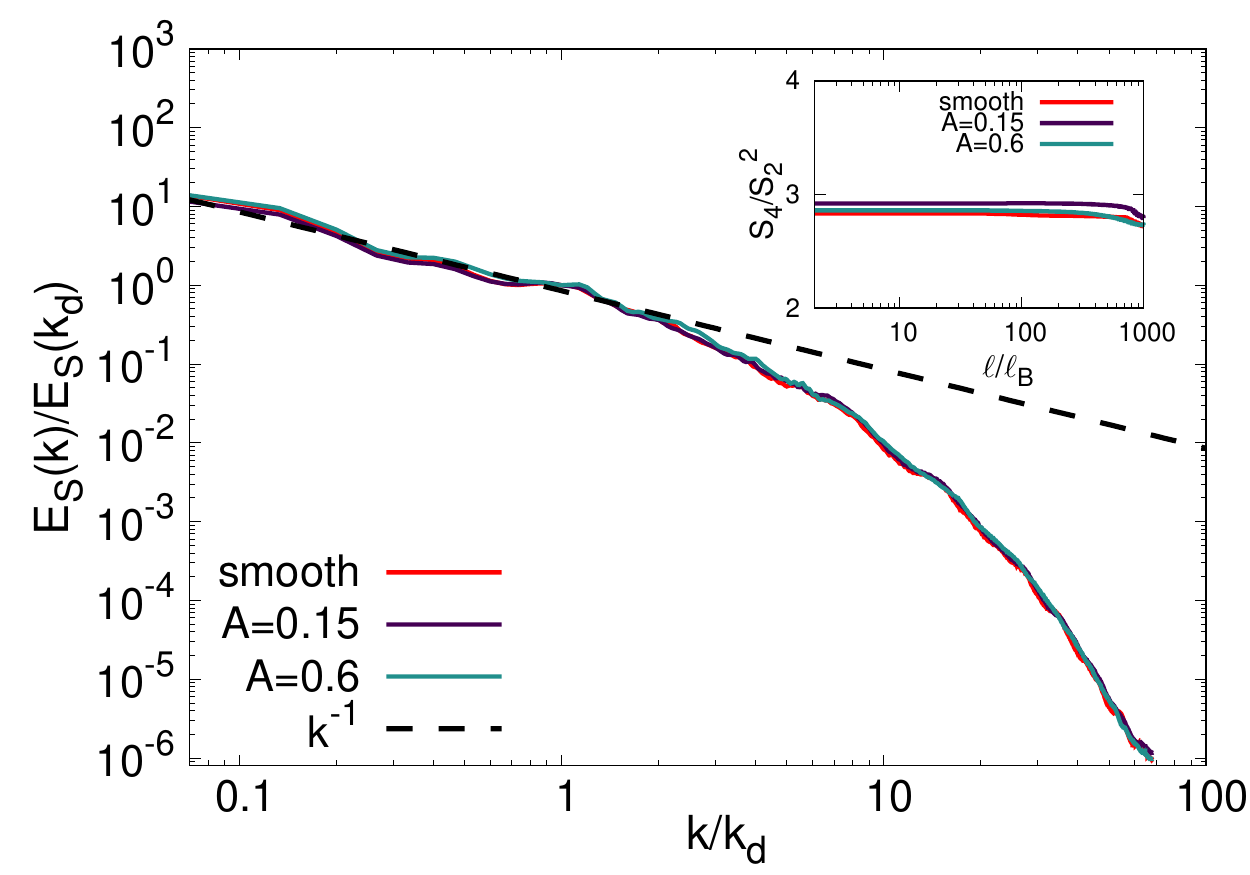}
\caption{(b)}
\label{fig:shapeb}
\end{subfigure}%
\caption{(a) Snapshot of the phytoplankton density field at time $t = 190$ (in non-dimensional units) for the simulation with $A=0.15$ and $Re=400$.
(b) Spectra of phytoplankton fluctuations $E_S(k)$ for the smooth obstacle and the two rough ones, normalized by 
$E_S(k_d)$. The spectra are computed along the $y$-direction and averaged over $1.5d \leq x \leq 10d$ and time. 
The inset shows the flatness $S_4/S_2^2$ of the $P$ distribution, versus the distance $\ell$ (normalized by the Batchelor scale), 
for the same three obstacle shapes. 
}
\end{figure}

We observe that the spatially averaged value of population density is very weakly dependent on the roughness of the island, with relative variations of the order of $3$-$5\%$ (not shown) for the cases here explored. This supports the fact that the global excitation of the system is independent of the specific characteristics of boundary layers. 
Also from the spectra of phytoplankton fluctuations, no appreciable differences could be detected. As shown in Fig.~\ref{fig:shapeb}, the spectra of $P$ fluctuations for the three cases are practically identical over a broad range of scales. While the boundary layer properties are affected, let us note that the same robustness of spectral properties was found also for the fluid velocity fluctuations further away from the obstacle. 
To further probe possible effects of the roughness on higher-order statistics, we also computed the flatness, defined as $S_4/S_2^2$, 
where $S_2$ and $S_4$ respectively are the $2^\mathrm{nd}$ and $4^\mathrm{th}$ order transversal structure functions of the scalar field $P(\bm{x},t)$. Independent of the details of the obstacle shape, we always found $S_4/S_2^2 \simeq 3$ (see inset of Fig.~\ref{fig:shapeb}), as for a Gaussian distribution, 
in the scale range $\ell/\ell_B \in [1,1000]$ (with $\ell_B$ the Batchelor scale). 
These results then indicate that the obstacle roughness does not have a significant impact on the global and local properties 
of the advected planktonic species (at $Re=400$).
Such an evidence, along with the weak dependence on the Reynolds number documented in the previous sections, suggests that even at higher values of $Re$ the roughness of the obstacle should not give rise to major modifications of the population dynamics.

\subsection{Impact of the coarse-graining}
\label{sec:subgrid}

In this last section, we report some results about a comparison between a fully resolved DNS, as used in the previous sections, 
and two coarse-grained simulations.
In this case, the flow and geometrical parameters are the same, but the grid is not sufficiently refined to resolve all the scales in the flow and scalars dynamics. This means that the simulations are under-resolved, and some small-scale effects are lacking.
This approach is related to the Large-eddy-simulation one \cite{pope}, where the unresolved scales are called sub-grid scales, and their effect should be reconstructed via some model. Our coarse-grained simulations represent the simplest case in which no sub-grid model is taken into account.
Such a comparison is important since DNS cannot be used to simulate realistic configurations where only a coarse-grained approach is computationally feasible.

\begin{figure}[ht]
\captionsetup[subfigure]{labelformat=empty,justification=centering}
\begin{subfigure}{.5\textwidth}
\centering
\includegraphics[width=0.8\textwidth]{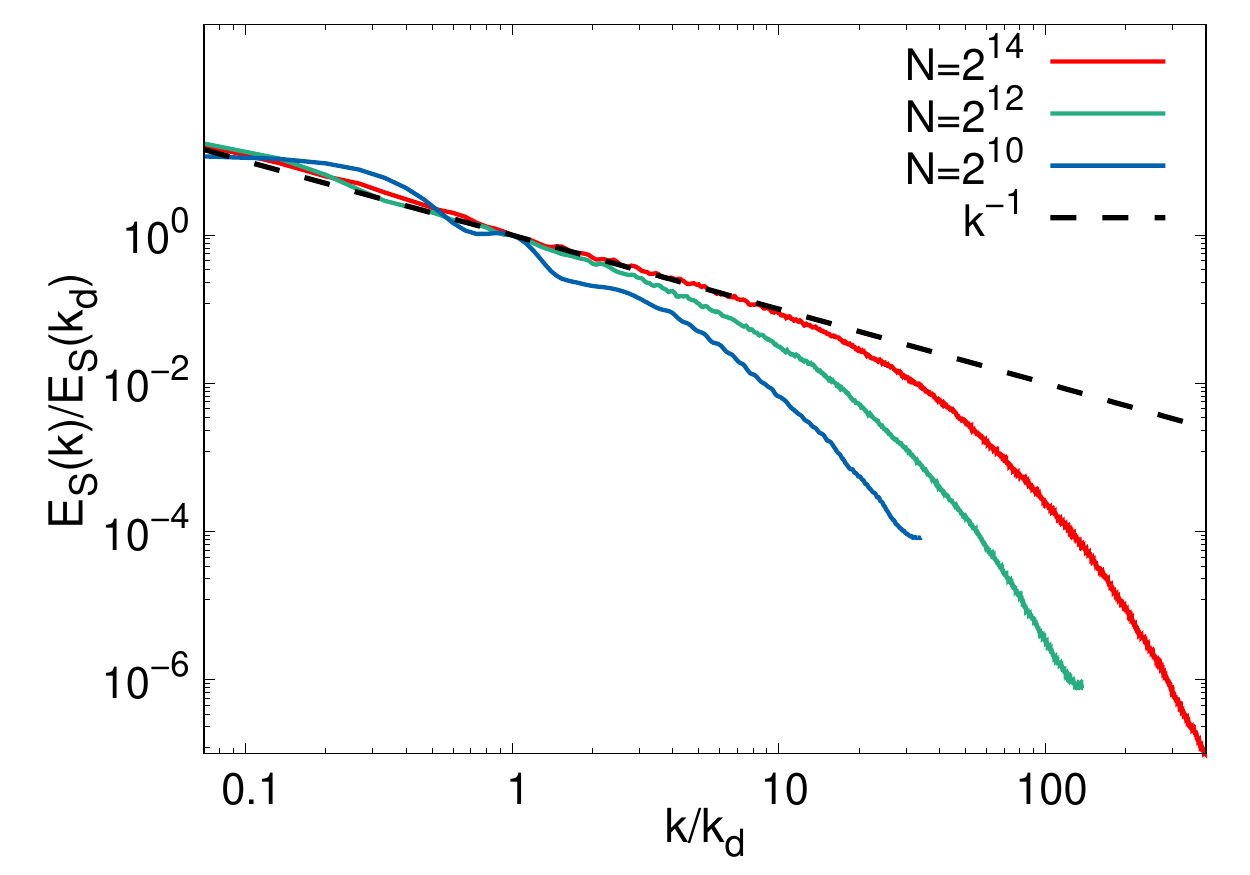}
\caption{(a)}
\label{fig:lesa}
\end{subfigure}%
\begin{subfigure}{.5\textwidth}
\centering
\includegraphics[width=0.8\textwidth]{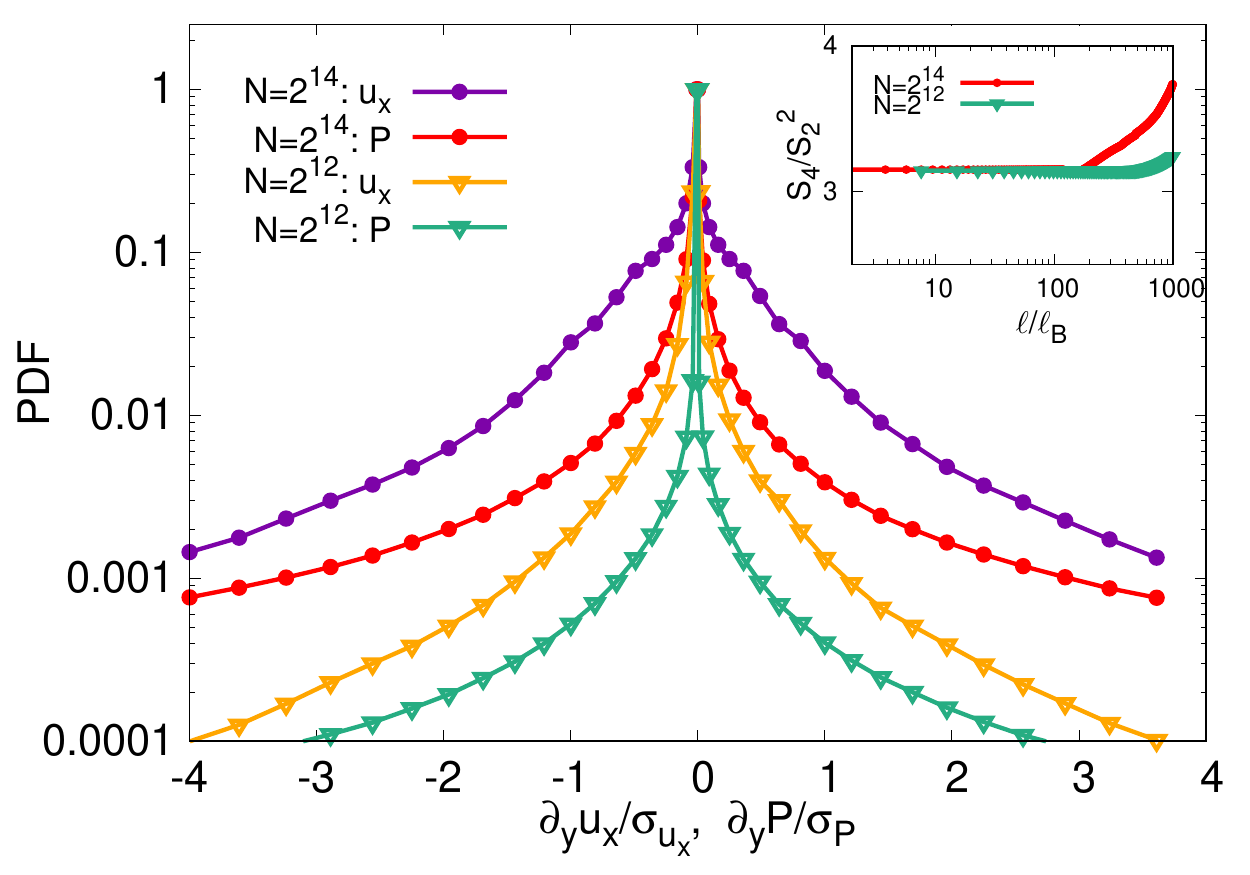}
\caption{(b)}
\label{fig:lesb}
\end{subfigure}%
\caption{(a) Spectra of phytoplankton fluctuations $E_S(k)$ at $Re = 20000$ at different maximum grid refinement levels $N$. 
(b) PDFs of the gradients, in the transversal direction, of longitudinal velocity 
and phytoplankton density for the simulation with $N=2^{14}$ (DNS) and with $N=2^{12}$ (LES) at $Re=20000$. 
All the data here correspond to centered fluctuations, rescaled by the corresponding standard deviation. In the inset the flatness of the phytoplankton density field versus the separation in the transversal direction
for the two simulations is shown.}
\end{figure}

We consider the most turbulent case at $Re=20000$ and a circular smooth obstacle. 
The numerical setup is identical in all cases, except for the maximum level of grid refinement: while for the DNS this is $N=2^{14}$ and the smallest resolved scale is $\Delta x \sim 1.5 \, \ell_B$, for the LES we have $N=2^{12}$ and $\Delta x \sim 6 \, \ell_B$ in the first case and $N=2^{10}$ and $\Delta x \sim 25 \, \ell_B$ in the second one. 
Concerning the global value of population density (results not shown), we can remark that while the case at $N=2^{12}$ still captures the correct behaviour over time, as predicted by the fully-resolved simulation, when the refinement level is reduced to $N=2^{10}$ important errors emerge, in terms of the average value $\overline{\langle P\rangle}$, as well as oscillations around it, generating a completely misleading picture of the dynamics (both the flow and the scalars appear more regular in time and space, resembling a case at lower $Re$).
This picture is confirmed by Fig.~\ref{fig:lesa}, which shows the spectra of phytoplankton fluctuations in the three cases: the spectra at $N=2^{14}$ and $N=2^{12}$ are quite similar, except for the decade of the smallest spatial scales, which cannot be captured by the under-resolved simulation; at $N=2^{10}$, instead, the spectrum differs from those at higher resolution also at large scales, confirming the inadequacy of this simulation to account for the dynamics at such high $Re$ number.
In Fig.~\ref{fig:lesb} we show the PDFs of the transversal gradients of the phytoplankton density, representative 
of the reactive scalar small-scale features, and of longitudinal velocity, for the cases at $N=2^{14}$ and $N=2^{12}$.
A first remark is that for both the DNS and the LES the $P$ gradients' PDF follows that of the fluid velocity, though with lower tails. 
More interestingly, it is also apparent that in the LES case the statistics of large-deviation events are 
less important, for both $\partial_y P$ and $\partial_y u_x$, so that the PDFs of the latter two fields are narrower and with faster decreasing tails, with respect to the DNS case. 
Similarly to the analysis of Sec.~\ref{sec:shape}, in the inset of Fig.~\ref{fig:lesb} we  present the flatness of the phytoplankton density field, considering increments 
in the $y$ direction. 
Clearly, for the LES case with $N=2^{12}$, the smallest separations that is possible to sample are larger than for the DNS case 
with $N=2^{14}$. Beyond this, the results indicate that for $\ell/\ell_B \geq 100$ the flatness starts to deviate from the value $3$ and in the case of the fully resolved simulation this feature is more evident. 
Our analysis suggests that, although a fully resolved simulation allows to catch all the statistical details of the dynamics, 
the smallest scales do not seem to have a significant impact on the overall dynamics. Consequently a coarse-grained approach, as in the case with $N=2^{12}$, appears valid if one is interested in large-scale dynamics (in particular it is important to note that with such resolution all the flow scales are resolved but this is not the case for the scalars, because 
$Sc \neq 1$). 
Instead, a very under-resolved approach ($N=2^{10}$), both in terms of flow and scalar scales, is misleading and incapable of capturing the correct large and small scale dynamics. 
Yet, it is worth emphasizing here that all these simulations provide the same blooming conditions.

\section{Conclusions}
\label{sec:conclu}

We have investigated predator-prey plankton dynamics in turbulent flows occurring in the wake of an obstacle, 
which is an idealized configuration intended to mimic an island in the ocean. 
Our purpose was to understand which hydrodynamic conditions trigger an algal bloom, and how active small scales may affect 
it.

Aiming to focus on the role played by the flow field, we chose to limit the complexity of the biological dynamics.
In this spirit we adopted the PZ model, perhaps the simplest multi-species model system allowing to reproduce the main features 
of algal blooms~\cite{tru}.
Since nutrients are not included, and the 
system is excitable, whenever the flow conditions are capable to trigger a bloom, 
these model dynamics will sustain it.
Furthermore, to evaluate the possible effect of the Reynolds number, we have performed three simulations up to $Re=20000$.

The comparison with previous studies  
in open flows obtained from kinematic models~\cite{neu,lo,chaos,NLHP2002,hl,Sandulescu_etal2008,BF2010,GF2020} has been instructive, and one of our main results has been to indicate that the key mechanism underlying the blooming is the presence in the flow of at least a region where the 
biological species remain trapped for very long time. 
If this feature is encountered, basically the same bloom is triggered irrespective of the precise features of the flow and 
of the Reynolds number. 
In fact, as in the case of PZ dynamics in a perturbed-jet kinematic flow~\cite{hl}, in our dynamical turbulent wake, advection
needs to be slow enough with respect to the phytoplankton growth, for this to occur.
Interestingly, we have been able to detect also 
a second transition, to de-excitation, for very slow advection, which could not be observed in that study 
but was found in another kinematic flow~\cite{NLHP2002}. 
Indeed, in our system an optimal value of order $1$ 
exists for the ratio between advection and biological time scales, which promotes maximum primary production.

We have further found that the presence of a chaotic saddle, whose role had been put in evidence
in previous kinematic studies, is not necessarily crucial for  
blooming, whereas for transport and mixing the details of the fluid dynamics are important. 
Indeed, we have verified that the plankton bloom occurs even in the laminar regime and this result allows us to estimate in a simple way the average biomass of the system, which in 
this case is concentrated in two straight filaments departing from the obstacle, approximately in the streamwise direction, at symmetric positions with respect to the centerline 
in the cross-stream direction. 
According to the model developed in~\cite{martin2000filament} for a single reactive tracer in a purely elongational flow, 
the width $\ell_f$ of a plankton filament is independent of the biological growth rate and solely determined by the physical parameters of the flow, 
$\ell_f \sim \sqrt{D/s}$, where $D$ is the effective diffusivity and $s$ the strain rate. In our case, the latter can be dimensionally estimated as $s \sim u_0/\delta_{BL}$, where 
$\delta_{BL} \sim \sqrt{\nu l_0/u_0}$ is the boundary-layer thickness, 
which gives $\ell_f \sim (D^2 \nu l_0/u_0^3)^{1/4}$. 
In this way, provided that a patch of population density equal to $P_a$ enters the domain, which is initially at $P_{eq}$, 
it is reasonable to expect that the minimum total 
phytoplankton biomass is given by $P_{tot} \sim 2P_a\ell_f L_f + P_{eq}L^2$ where $L_f$ is the length of 
each filament and $L$ the linear domain size. 

From this, using our parameter values, the average population density, normalized by its equilibrium value, 
is predicted to be $\overline{\langle P\rangle} /P_{eq} \sim 1.15$, 
not far from the numerical value of $1.75$ that we find at $Re=10$. The discrepancy is grasped by considering
the reactive nature of phytoplankton, which on average grows in a blooming situation, while being stretched by the flow, before the zooplankton can effectively graze on it. 

We have then studied the phenomenology of the small scales at high $Re$.
While the large-scale dynamics, and with it the main characteristics of the bloom, remain basically unchanged, similarly to what reported in 
previous studies~\cite{neufeld,chaos}, a smooth-filamental transition takes place, depending on the 
relative importance of the strain-rate intensity and the biological growth rate. 
This point was studied by varying the turbulent intensity in our simulations, 
instead of the biological parameters as in the case of kinematic flows, which seems to us more interesting 
in relation to realistic situations. We found that the transition, manifesting as the appearance of fractal 
features in transects of the $P$ field, occurs at  large Reynolds numbers ($Re>2000$), 
when more small scales are present in the velocity field.  
This suggests that small-scale fluid motions locally affect the fine-scale spatial distribution 
of the planktonic species, if the flow is turbulent enough. 
We have also addressed the problem of plankton patchiness by analysing the spectral properties of 
$P$ and $Z$ fields, as well as the correlation between the spatial structures of the latter and those 
of the flow field. Our finding is a neat $k^{-1}$ scaling, over more than two decades, for the 
variance spectra of the population density fields.
This provides a clear indication that reactive scalars are not different from passive 
(non-reactive) ones with regard to the statistical properties, and lends support to the arguments developed in 
the framework of simplified theories for interacting species in 2D turbulent flows~\cite{powell}.
That represents also an important result of the present work. 

As for the correlation between flow structures and population patchiness, the results 
previously reported, from both idealized~\cite{hl,Sandulescu_etal2007,Sandulescu_etal2008} 
and realistic models~\cite{Carrasco_etal_2014,Levy_etal_2018}, 
or from observations~\cite{McGillicuddy_2016,Levy_etal_2018,Zhang_etal_2019}, are varied, 
due to the relevance of different mechanisms and the variation of their relative weight in different regions.
The results from our simulations provide clear evidence that, in the present setup, phytoplankton mainly concentrates in filamentary structures, winding in the periphery of vortices.
At the same time, the biological interactions control the relative abundance of zooplankton and phytoplankton, locally. 
We hope that such outcome can contribute to shed light on the complex organization of plankton
with respect to the characteristics of the carrying flow, beyond the peculiarities of the flow and biological dynamics 
here considered. Indeed, some similarities with the present phenomenology have been observed, under some circumstances, also using more
realistic biological models~\cite{Sandulescu_etal2007,Sandulescu_etal2008,GF2020}. 

We have furthermore verified that all these results are \emph{quantitatively} independent from  the roughness of the obstacle.
For this purpose, we considered islands with different fractal contours~\cite{rough}, as representative of rocky coastlines~\cite{BCDS2008}.
This result indicates that the detailed spatial structure of boundary conditions has, to good extent, 
a minor role and corroborates the findings highlighting that the mechanism controlling blooms is mainly related to the presence of flow regions trapping the planktonic species. 

The last main result of this work concerns the possibility to use a coarse-grained approach instead of resolving all the dynamical scales, as here, which is impossible in a realistic situation. 
We have analyzed simulations with progressively coarser spatial resolution and our results point out two possibilities 
with respect to the level of information sought: 
(i) the dynamics is not very sensitive to small scales, and therefore accurate statistical properties at large 
and intermediate scales can be obtained with a coarser approach, when all these scales are resolved (a well-resolved LES, in the computational-fluid-dynamics jargon); 
(ii) if the focus is only on very large-scale features,  for instance to answer the question \emph{is there a bloom or not?}, a large-scale approach seems feasible, with the caveat that
the statistical properties cannot be well reproduced.

\section{Acknowledgements}
\noindent This work was granted access to the HPC resources [MESU] of the HPCaVe centre at UPMC-Sorbonne University. 
We thank St\'ephane Popinet for his precious suggestions.

\appendix 
\section{Numerical method}
\label{app:num}

Basilisk is an open source code written using an extension to the C programming language, called Basilisk C, for the resolution of partial differential equations (see \url{http://basilisk.fr}) . 
Space is discretized using a Cartesian tree-based grid, which is adaptively varied according to the $Re$ and $Sc$ numbers for well-resolving all the scales.
Two primary criteria are used to decide where to refine the mesh. They are based on a wavelet-decomposition of the velocity and scalars and volume fraction fields respectively~\cite{van}. The velocity and scalars criterion is mostly sensitive to the second-derivative of the fields and guarantees refinement in developing boundary layers and wakes. The volume fraction criterion is sensitive to the curvature of the interface and guarantees the accurate description of the shape of the obstacle.
Both criteria are usually combined with a maximum allowed level of refinement.
Boundaries of general shape are reconstructed using an integral (i.e. finite volume) formulation which takes into account the volume and area fractions of intersection of the embedded boundary with the mesh~\cite{Johansen1998ACG}.
The numerical scheme implemented in Basilisk is 
described, {\it e.g.}, in~\cite{pop2009}. 
The Navier–Stokes equations are integrated by a projection method~\cite{cho}. Standard second-order numerical schemes for the spatial gradients 
are used~\cite{pop2003,pop2009,lag}. In particular, the velocity advection term $\partial_j(u_j u_i)^{n+1/2}$ is estimated by means of the 
Bell-Colella-Glaz second/third-order unsplit upwind scheme~\cite{pop2003}. In this way, the problem is reduced to the solution of a
Helmholtz–Poisson problem for each primitive variable and a Poisson problem for the pressure correction terms. 
Both the Helmholtz-Poisson and Poisson problems are solved using an efficient multilevel solver~\cite{pop2003,pop2015}.
The time advancing is made through a fractional-step method using a staggered discretization in time of the velocity and 
the scalar fields~\cite{pop2009}: one supposes the velocity field to be known at time $n$ and the scalar fields (pressure, density, plankton) 
to be known at time $n -1/2$ and one computes velocity at time $n+1$ and scalars at time $n+1/2$.\\
In all the simulations, we adopted the following boundary conditions for the flow:

\begin{subequations}
\begin{align}
&[u_n]_{left} = u_0\\
&[u_t]_{left} = 0\\
&[\partial_n u_n]_{right} = 0\\
&[u_n]_{top} = [u_n]_{bottom}= 0\\
&[\partial_n u_t]_{top} = [\partial_n u_t]_{bottom}= 0\\
&[u_n]_{obstacle} = [u_t]_{obstacle} = 0
\end{align}
\end{subequations}

i.e. an inflow/outflow condition is imposed on the left/right side of the domain, while free-slip conditions hold at the boundaries in the $y$-direction and a no-slip condition on the obstacle. The subscripts $n$ and $t$ stand for the normal and tangential component relative to the boundary walls.
In order to omit issues with inconsistent boundary conditions, we introduced a damping layer near the in and out flow boundaries.\\
\noindent As regarding the two scalars, they are kept at the equilibrium values ($P_{eq},Z_{eq}$) at all sides of the domain, while on the obstacle a no-flux condition is imposed:

\begin{subequations}
\begin{align}
&[\partial_n P]_{obstacle} = 0,\\
&[\partial_n Z]_{obstacle} = 0.
\end{align}
\end{subequations}

\section{Comparison with a passive scalar}
\label{app:comp}

\begin{figure}[ht]
	\centering
	\includegraphics[width=0.475\textwidth]{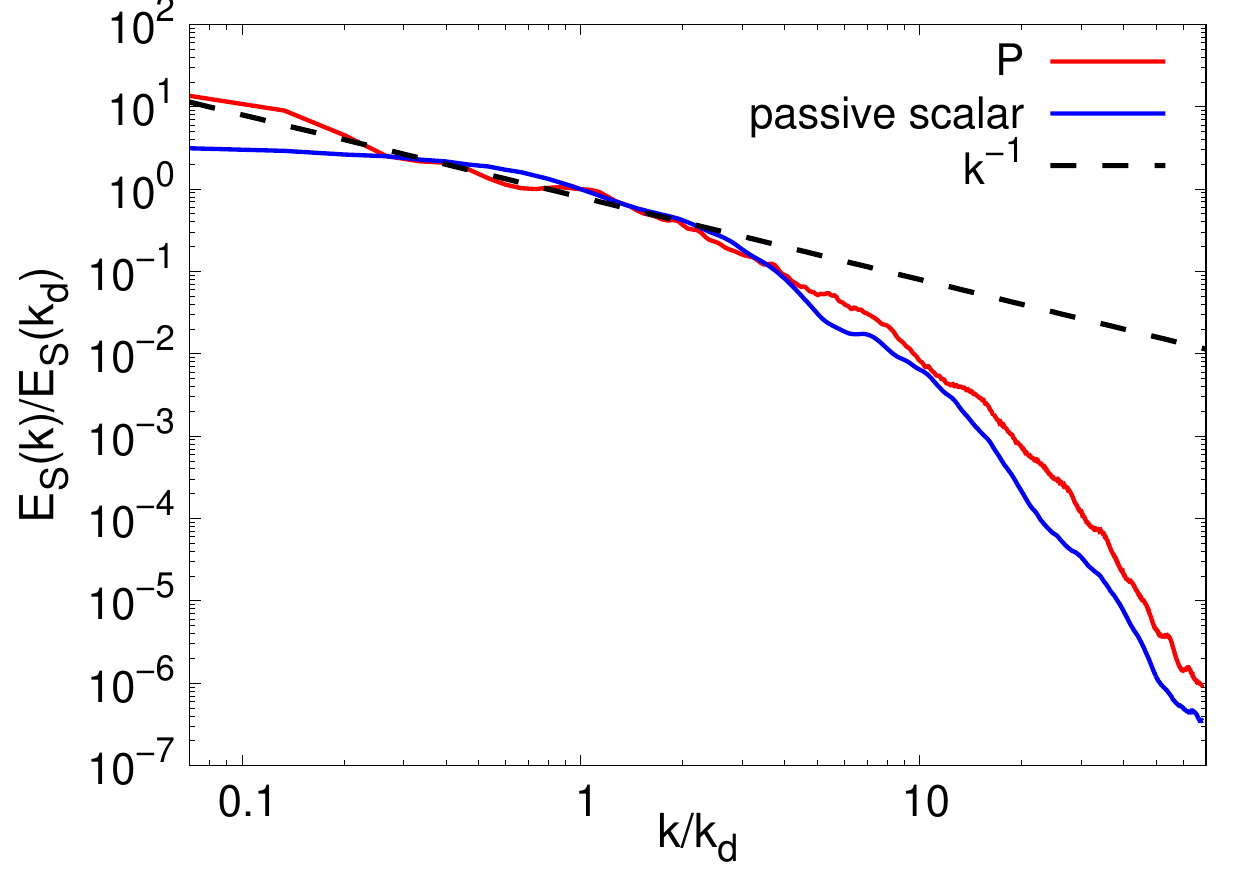}
	\caption{Spectra of phytoplankton density, and non-reactive passive scalar, fluctuations. 
        The spectra are computed along the $y$-direction 
        and averaged over $1.5d \leq x \leq 10d$ and time.}
	\label{fig:spe2}
\end{figure}

In order to better understand the effect of the growth dynamics, and of the biological interactions, on plankton spectra, 
we performed a simulation at $Re=400$ (with a smooth circular obstacle) considering a passive, non-reactive, scalar, 
obeying the same equations of the plankton species, excepting for the reaction term which is in this case equal to $0$.
The choice of this particular value of $Re$ was motivated by the low computational cost. 
The boundary conditions are the same adopted for the phytoplankton, excepting for the condition on the obstacle where we impose a Dirichlet boundary condition by choosing a constant value higher than the surroundings (we also verified that the exact numerical value has no impact on the dynamics).
Our focus is here on possible differences induced by the reactive evolution in comparison to the passive one.
We expected 
that these differences, if they exist, could be visible at large scales where the biological activity is predominant with respect 
to the turbulent environmental effects~\cite{DP1976,powell}. In Fig.~\ref{fig:spe2}, the spectra of the phytoplankton 
density, as well as of the non-reactive tracer, fluctuations at $Re=400$ are shown. 
In both cases, the spectrum is compatible with a scaling $\sim k^{-1}$ 
in the enstrophy inertial range, then a steeper slope typical of a viscous range is found. 
Generally, the plankton spectrum appears to follow the $-1$ slope at scales larger than the injection scale, 
where the passive scalar spectrum is flatter. 

\section{Dynamic visualizations}
\label{app:video}

To further illustrate the dynamics of phytoplankton, two animations of the population density are provided at $Re = 400$ (see video1.mp4) and $Re = 20000$ (see video2.mp4).
In both cases, the phytoplankton starts to grow in the boundary layer around the obstacle: in the first case, it oscillates regularly in time and symmetrically with respect to the centerline 
in the cross-stream direction, mainly winding around the vortex cores. 
In the second animation (video2.mp4), the boundary layer where the scalar accumulates is not visible anymore because of its small thickness; the phytoplankton distribution is now highly irregular in time and space: it accelerates downstream of the obstacle and organizes in filaments which are thinner with respect to the previous case, as expected considering
that the filament width has been found to be a function of both the diffusion coefficient and strain rate of the flow (see the discussion in the Conclusions of the main text). 




\begin{thebibliography}{69}%
\makeatletter
\providecommand \@ifxundefined [1]{%
 \@ifx{#1\undefined}}%
\providecommand \@ifnum [1]{%
 \ifnum #1\expandafter \@firstoftwo
 \else \expandafter \@secondoftwo
 \fi}%
\providecommand \@ifx [1]{%
 \ifx #1\expandafter \@firstoftwo
 \else \expandafter \@secondoftwo
 \fi}%
\providecommand \natexlab [1]{#1}%
\providecommand \enquote  [1]{``#1''}%
\providecommand \bibnamefont  [1]{#1}%
\providecommand \bibfnamefont [1]{#1}%
\providecommand \citenamefont [1]{#1}%
\providecommand \href@noop [0]{\@secondoftwo}%
\providecommand \href [0]{\begingroup \@sanitize@url \@href}%
\providecommand \@href[1]{\@@startlink{#1}\@@href}%
\providecommand \@@href[1]{\endgroup#1\@@endlink}%
\providecommand \@sanitize@url [0]{\catcode `\\12\catcode `\$12\catcode
  `\&12\catcode `\#12\catcode `\^12\catcode `\_12\catcode `\%12\relax}%
\providecommand \@@startlink[1]{}%
\providecommand \@@endlink[0]{}%
\providecommand \url  [0]{\begingroup\@sanitize@url \@url }%
\providecommand \@url [1]{\endgroup\@href {#1}{\urlprefix }}%
\providecommand \urlprefix  [0]{URL }%
\providecommand \Eprint [0]{\href }%
\providecommand \doibase [0]{https://doi.org/}%
\providecommand \selectlanguage [0]{\@gobble}%
\providecommand \bibinfo  [0]{\@secondoftwo}%
\providecommand \bibfield  [0]{\@secondoftwo}%
\providecommand \translation [1]{[#1]}%
\providecommand \BibitemOpen [0]{}%
\providecommand \bibitemStop [0]{}%
\providecommand \bibitemNoStop [0]{.\EOS\space}%
\providecommand \EOS [0]{\spacefactor3000\relax}%
\providecommand \BibitemShut  [1]{\csname bibitem#1\endcsname}%
\let\auto@bib@innerbib\@empty

\bibitem [{\citenamefont {Mann}\ and\ \citenamefont {Lazier}(2005)}]{ML2005}%
  \BibitemOpen
  \bibfield  {author} {\bibinfo {author} {\bibfnamefont {K.}~\bibnamefont
  {Mann}}\ and\ \bibinfo {author} {\bibfnamefont {J.}~\bibnamefont {Lazier}},\
  }\href@noop {} {\emph {\bibinfo {title} {Dynamics of marine ecosystems:
  biological-physical interactions in the oceans}}}\ (\bibinfo  {publisher}
  {New York, NY: Wiley},\ \bibinfo {year} {2005})\BibitemShut {NoStop}%
\bibitem [{\citenamefont {Williams}\ and\ \citenamefont
  {Follows}(2011)}]{WF2011}%
  \BibitemOpen
  \bibfield  {author} {\bibinfo {author} {\bibfnamefont {R.~G.}\ \bibnamefont
  {Williams}}\ and\ \bibinfo {author} {\bibfnamefont {M.~J.}\ \bibnamefont
  {Follows}},\ }\href@noop {} {\emph {\bibinfo {title} {Ocean dynamics and the
  carbon cycle: principles and mechanisms}}}\ (\bibinfo  {publisher}
  {Cambridge, UK: Cambridge University Press},\ \bibinfo {year}
  {2011})\BibitemShut {NoStop}%
\bibitem [{\citenamefont {Sellner}\ \emph {et~al.}(2003)\citenamefont
  {Sellner}, \citenamefont {Doucette},\ and\ \citenamefont
  {Kirkpatrick}}]{Sellner_etal_2003}%
  \BibitemOpen
  \bibfield  {author} {\bibinfo {author} {\bibfnamefont {K.~G.}\ \bibnamefont
  {Sellner}}, \bibinfo {author} {\bibfnamefont {G.~J.}\ \bibnamefont
  {Doucette}},\ and\ \bibinfo {author} {\bibfnamefont {G.~J.}\ \bibnamefont
  {Kirkpatrick}},\ }\bibfield  {title} {\bibinfo {title} {Harmful algal blooms:
  causes, impacts and detection},\ }\href@noop {} {\bibfield  {journal}
  {\bibinfo  {journal} {J. Ind. Microbiol. Biotechnol.}\ }\textbf {\bibinfo
  {volume} {30}},\ \bibinfo {pages} {383} (\bibinfo {year} {2003})}\BibitemShut
  {NoStop}%
\bibitem [{\citenamefont {Kahru}\ and\ \citenamefont
  {Mitchell}(2008)}]{KM2008}%
  \BibitemOpen
  \bibfield  {author} {\bibinfo {author} {\bibfnamefont {M.}~\bibnamefont
  {Kahru}}\ and\ \bibinfo {author} {\bibfnamefont {B.~G.}\ \bibnamefont
  {Mitchell}},\ }\bibfield  {title} {\bibinfo {title} {Ocean color reveals
  increased blooms in various parts of the world},\ }\href@noop {} {\bibfield
  {journal} {\bibinfo  {journal} {Eos}\ }\textbf {\bibinfo {volume} {89}},\
  \bibinfo {pages} {170} (\bibinfo {year} {2008})}\BibitemShut {NoStop}%
\bibitem [{\citenamefont {Guseva}\ and\ \citenamefont {Feudel}(2020)}]{GF2020}%
  \BibitemOpen
  \bibfield  {author} {\bibinfo {author} {\bibfnamefont {K.}~\bibnamefont
  {Guseva}}\ and\ \bibinfo {author} {\bibfnamefont {U.}~\bibnamefont
  {Feudel}},\ }\bibfield  {title} {\bibinfo {title} {Numerical modelling of the
  effect of intermittent upwelling events on plankton blooms},\ }\href@noop {}
  {\bibfield  {journal} {\bibinfo  {journal} {J. R. Soc. Interface}\ }\textbf
  {\bibinfo {volume} {17}},\ \bibinfo {pages} {20190889} (\bibinfo {year}
  {2020})}\BibitemShut {NoStop}%
\bibitem [{\citenamefont {Denman}\ and\ \citenamefont
  {Gargett}(1995)}]{denman1995biological}%
  \BibitemOpen
  \bibfield  {author} {\bibinfo {author} {\bibfnamefont {K.}~\bibnamefont
  {Denman}}\ and\ \bibinfo {author} {\bibfnamefont {A.}~\bibnamefont
  {Gargett}},\ }\bibfield  {title} {\bibinfo {title} {Biological-physical
  interactions in the upper ocean: the role of vertical and small scale
  transport processes},\ }\href@noop {} {\bibfield  {journal} {\bibinfo
  {journal} {Oceanographic Literature Review}\ }\textbf {\bibinfo {volume}
  {8}},\ \bibinfo {pages} {662} (\bibinfo {year} {1995})}\BibitemShut {NoStop}%
\bibitem [{\citenamefont {Huisman}\ \emph {et~al.}(2002)\citenamefont
  {Huisman}, \citenamefont {Array\'as}, \citenamefont {Ebert},\ and\
  \citenamefont {Sommeijer}}]{Huisman_etal_2002}%
  \BibitemOpen
  \bibfield  {author} {\bibinfo {author} {\bibfnamefont {J.}~\bibnamefont
  {Huisman}}, \bibinfo {author} {\bibfnamefont {M.}~\bibnamefont {Array\'as}},
  \bibinfo {author} {\bibfnamefont {U.}~\bibnamefont {Ebert}},\ and\ \bibinfo
  {author} {\bibfnamefont {B.}~\bibnamefont {Sommeijer}},\ }\bibfield  {title}
  {\bibinfo {title} {How do sinking phytoplankton species manage to persist?},\
  }\href@noop {} {\bibfield  {journal} {\bibinfo  {journal} {Am. Nat.}\
  }\textbf {\bibinfo {volume} {159}},\ \bibinfo {pages} {245} (\bibinfo {year}
  {2002})}\BibitemShut {NoStop}%
\bibitem [{\citenamefont {Lindemann}\ \emph {et~al.}(2017)\citenamefont
  {Lindemann}, \citenamefont {Visser},\ and\ \citenamefont
  {Mariani}}]{Lindemann_etal_2017}%
  \BibitemOpen
  \bibfield  {author} {\bibinfo {author} {\bibfnamefont {C.}~\bibnamefont
  {Lindemann}}, \bibinfo {author} {\bibfnamefont {A.}~\bibnamefont {Visser}},\
  and\ \bibinfo {author} {\bibfnamefont {P.}~\bibnamefont {Mariani}},\
  }\bibfield  {title} {\bibinfo {title} {Dynamics of phytoplankton blooms in
  turbulent vortex cells},\ }\href@noop {} {\bibfield  {journal} {\bibinfo
  {journal} {J. R. Soc. Interface}\ }\textbf {\bibinfo {volume} {14}},\
  \bibinfo {pages} {20170453} (\bibinfo {year} {2017})}\BibitemShut {NoStop}%
\bibitem [{\citenamefont {Abraham}(1998)}]{abra}%
  \BibitemOpen
  \bibfield  {author} {\bibinfo {author} {\bibfnamefont {E.}~\bibnamefont
  {Abraham}},\ }\bibfield  {title} {\bibinfo {title} {The generation of
  plankton patchiness by turbulent stirring},\ }\href@noop {} {\bibfield
  {journal} {\bibinfo  {journal} {Nature}\ }\textbf {\bibinfo {volume} {391}},\
  \bibinfo {pages} {577} (\bibinfo {year} {1998})}\BibitemShut {NoStop}%
\bibitem [{\citenamefont {Martin}\ \emph {et~al.}(2002)\citenamefont {Martin},
  \citenamefont {Richards}, \citenamefont {Bracco},\ and\ \citenamefont
  {Provenzale}}]{martin2002}%
  \BibitemOpen
  \bibfield  {author} {\bibinfo {author} {\bibfnamefont {A.}~\bibnamefont
  {Martin}}, \bibinfo {author} {\bibfnamefont {K.}~\bibnamefont {Richards}},
  \bibinfo {author} {\bibfnamefont {A.}~\bibnamefont {Bracco}},\ and\ \bibinfo
  {author} {\bibfnamefont {A.}~\bibnamefont {Provenzale}},\ }\bibfield  {title}
  {\bibinfo {title} {Patchy productivity in the open ocean},\ }\href
  {https://doi.org/https://doi.org/10.1029/2001GB001449} {\bibfield  {journal}
  {\bibinfo  {journal} {Global Biogeochemical Cycles}\ }\textbf {\bibinfo
  {volume} {16}},\ \bibinfo {pages} {9} (\bibinfo {year} {2002})}\BibitemShut
  {NoStop}%
\bibitem [{\citenamefont {Martin}(2003)}]{martin2003phytoplankton}%
  \BibitemOpen
  \bibfield  {author} {\bibinfo {author} {\bibfnamefont {A.}~\bibnamefont
  {Martin}},\ }\bibfield  {title} {\bibinfo {title} {Phytoplankton patchiness:
  the role of lateral stirring and mixing},\ }\href@noop {} {\bibfield
  {journal} {\bibinfo  {journal} {Progress in oceanography}\ }\textbf {\bibinfo
  {volume} {57}},\ \bibinfo {pages} {125} (\bibinfo {year} {2003})}\BibitemShut
  {NoStop}%
\bibitem [{\citenamefont {R.}\ \emph {et~al.}(2003)\citenamefont {R.},
  \citenamefont {Hillary}, \citenamefont {Bees}, \citenamefont {Sancho},\ and\
  \citenamefont {Sagu{\'e}s}}]{Reigada_etal_2003}%
  \BibitemOpen
  \bibfield  {author} {\bibinfo {author} {\bibfnamefont {R.}~\bibnamefont
  {R.}}, \bibinfo {author} {\bibfnamefont {R.~M.}\ \bibnamefont {Hillary}},
  \bibinfo {author} {\bibfnamefont {M.~A.}\ \bibnamefont {Bees}}, \bibinfo
  {author} {\bibfnamefont {J.~M.}\ \bibnamefont {Sancho}},\ and\ \bibinfo
  {author} {\bibfnamefont {F.}~\bibnamefont {Sagu{\'e}s}},\ }\bibfield  {title}
  {\bibinfo {title} {Plankton blooms induced by turbulent flows},\ }\href@noop
  {} {\bibfield  {journal} {\bibinfo  {journal} {Proc. R. Soc. Lond. B}\
  }\textbf {\bibinfo {volume} {270}},\ \bibinfo {pages} {875} (\bibinfo {year}
  {2003})}\BibitemShut {NoStop}%
\bibitem [{\citenamefont {L{\'o}pez}\ \emph {et~al.}(2001)\citenamefont
  {L{\'o}pez}, \citenamefont {Neufeld}, \citenamefont
  {Hern{\'a}ndez-Garc{\'\i}a},\ and\ \citenamefont {Haynes}}]{lo}%
  \BibitemOpen
  \bibfield  {author} {\bibinfo {author} {\bibfnamefont {C.}~\bibnamefont
  {L{\'o}pez}}, \bibinfo {author} {\bibfnamefont {Z.}~\bibnamefont {Neufeld}},
  \bibinfo {author} {\bibfnamefont {E.}~\bibnamefont
  {Hern{\'a}ndez-Garc{\'\i}a}},\ and\ \bibinfo {author} {\bibfnamefont
  {P.}~\bibnamefont {Haynes}},\ }\bibfield  {title} {\bibinfo {title} {Chaotic
  advection of reacting substances: Plankton dynamics on a meandering jet},\
  }\href@noop {} {\bibfield  {journal} {\bibinfo  {journal} {Physics and
  Chemistry of the Earth, Part B: Hydrology, Oceans and Atmosphere}\ }\textbf
  {\bibinfo {volume} {26}},\ \bibinfo {pages} {313} (\bibinfo {year}
  {2001})}\BibitemShut {NoStop}%
\bibitem [{\citenamefont {Goodman}\ and\ \citenamefont
  {Robinson}(2008)}]{GR2008}%
  \BibitemOpen
  \bibfield  {author} {\bibinfo {author} {\bibfnamefont {L.}~\bibnamefont
  {Goodman}}\ and\ \bibinfo {author} {\bibfnamefont {A.~R.}\ \bibnamefont
  {Robinson}},\ }\bibfield  {title} {\bibinfo {title} {On the theory of
  advective effects on biological dynamics in the sea. iii. the role of
  turbulence in biological–physical interactions},\ }\href@noop {} {\bibfield
   {journal} {\bibinfo  {journal} {Proc. R. Soc. A}\ }\textbf {\bibinfo
  {volume} {464}},\ \bibinfo {pages} {555} (\bibinfo {year}
  {2008})}\BibitemShut {NoStop}%
\bibitem [{\citenamefont {L{\'e}vy}(2008)}]{Levy2008}%
  \BibitemOpen
  \bibfield  {author} {\bibinfo {author} {\bibfnamefont {M.}~\bibnamefont
  {L{\'e}vy}},\ }\bibfield  {title} {\bibinfo {title} {The modulation of
  biological production by oceanic mesoscale turbulence},\ }\href@noop {}
  {\bibfield  {journal} {\bibinfo  {journal} {Lect. Notes Phys.}\ }\textbf
  {\bibinfo {volume} {744}},\ \bibinfo {pages} {219} (\bibinfo {year}
  {2008})}\BibitemShut {NoStop}%
\bibitem [{\citenamefont {McGillicuddy}(2016)}]{McGillicuddy_2016}%
  \BibitemOpen
  \bibfield  {author} {\bibinfo {author} {\bibfnamefont {D.~J.~J.}\
  \bibnamefont {McGillicuddy}},\ }\bibfield  {title} {\bibinfo {title}
  {Mechanisms of physical-biological-biogeochemical interaction at the oceanic
  mesoscale},\ }\href@noop {} {\bibfield  {journal} {\bibinfo  {journal} {Annu.
  Rev. Mar. Sci.}\ }\textbf {\bibinfo {volume} {8}},\ \bibinfo {pages} {125}
  (\bibinfo {year} {2016})}\BibitemShut {NoStop}%
\bibitem [{\citenamefont {L{\'e}vy}\ \emph {et~al.}(2018)\citenamefont
  {L{\'e}vy}, \citenamefont {Franks},\ and\ \citenamefont
  {Smith}}]{Levy_etal_2018}%
  \BibitemOpen
  \bibfield  {author} {\bibinfo {author} {\bibfnamefont {M.}~\bibnamefont
  {L{\'e}vy}}, \bibinfo {author} {\bibfnamefont {P.~J.~S.}\ \bibnamefont
  {Franks}},\ and\ \bibinfo {author} {\bibfnamefont {K.~S.}\ \bibnamefont
  {Smith}},\ }\bibfield  {title} {\bibinfo {title} {The role of submesoscale
  currents in structuring marine ecosystems},\ }\href@noop {} {\bibfield
  {journal} {\bibinfo  {journal} {Nat. Commun.}\ }\textbf {\bibinfo {volume}
  {9}},\ \bibinfo {pages} {4758} (\bibinfo {year} {2018})}\BibitemShut
  {NoStop}%
\bibitem [{\citenamefont {Zhang}\ \emph {et~al.}(2019)\citenamefont {Zhang},
  \citenamefont {Qiu}, \citenamefont {Klein},\ and\ \citenamefont
  {Travis}}]{Zhang_etal_2019}%
  \BibitemOpen
  \bibfield  {author} {\bibinfo {author} {\bibfnamefont {Z.}~\bibnamefont
  {Zhang}}, \bibinfo {author} {\bibfnamefont {B.}~\bibnamefont {Qiu}}, \bibinfo
  {author} {\bibfnamefont {P.}~\bibnamefont {Klein}},\ and\ \bibinfo {author}
  {\bibfnamefont {S.}~\bibnamefont {Travis}},\ }\bibfield  {title} {\bibinfo
  {title} {The influence of geostrophic strain on oceanic ageostrophic motion
  and surface chlorophyll},\ }\href@noop {} {\bibfield  {journal} {\bibinfo
  {journal} {Nat. Commun.}\ }\textbf {\bibinfo {volume} {10}},\ \bibinfo
  {pages} {2838} (\bibinfo {year} {2019})}\BibitemShut {NoStop}%
\bibitem [{\citenamefont {Hern{\'a}ndez-Garc{\'\i}a}\ and\ \citenamefont
  {L{\'o}pez}(2004)}]{hl}%
  \BibitemOpen
  \bibfield  {author} {\bibinfo {author} {\bibfnamefont {E.}~\bibnamefont
  {Hern{\'a}ndez-Garc{\'\i}a}}\ and\ \bibinfo {author} {\bibfnamefont
  {C.}~\bibnamefont {L{\'o}pez}},\ }\bibfield  {title} {\bibinfo {title}
  {Sustained plankton blooms under open chaotic flows},\ }\href@noop {}
  {\bibfield  {journal} {\bibinfo  {journal} {Ecological Complexity}\ }\textbf
  {\bibinfo {volume} {1}},\ \bibinfo {pages} {253} (\bibinfo {year}
  {2004})}\BibitemShut {NoStop}%
\bibitem [{\citenamefont {Denman}\ and\ \citenamefont {Platt}(1976)}]{DP1976}%
  \BibitemOpen
  \bibfield  {author} {\bibinfo {author} {\bibfnamefont {K.}~\bibnamefont
  {Denman}}\ and\ \bibinfo {author} {\bibfnamefont {T.}~\bibnamefont {Platt}},\
  }\bibfield  {title} {\bibinfo {title} {The variance spectrum of phytoplankton
  in a turbulent ocean},\ }\href@noop {} {\bibfield  {journal} {\bibinfo
  {journal} {J. mar. Res}\ }\textbf {\bibinfo {volume} {34}},\ \bibinfo {pages}
  {593} (\bibinfo {year} {1976})}\BibitemShut {NoStop}%
\bibitem [{\citenamefont {Smith}\ \emph {et~al.}(1988)\citenamefont {Smith},
  \citenamefont {Zhang},\ and\ \citenamefont {Michaelsen}}]{Smith_etal_1988}%
  \BibitemOpen
  \bibfield  {author} {\bibinfo {author} {\bibfnamefont {R.~C.}\ \bibnamefont
  {Smith}}, \bibinfo {author} {\bibfnamefont {X.}~\bibnamefont {Zhang}},\ and\
  \bibinfo {author} {\bibfnamefont {J.}~\bibnamefont {Michaelsen}},\ }\bibfield
   {title} {\bibinfo {title} {Variability of pigment biomass in the
  {C}alifornia current system as determined by satellite imagery. 1. spatial
  variability},\ }\href@noop {} {\bibfield  {journal} {\bibinfo  {journal} {J.
  Geophys. Res.}\ }\textbf {\bibinfo {volume} {93}},\ \bibinfo {pages} {10863}
  (\bibinfo {year} {1988})}\BibitemShut {NoStop}%
\bibitem [{\citenamefont {Martin}\ and\ \citenamefont
  {Srokosz}(2002)}]{MS2002}%
  \BibitemOpen
  \bibfield  {author} {\bibinfo {author} {\bibfnamefont {A.~P.}\ \bibnamefont
  {Martin}}\ and\ \bibinfo {author} {\bibfnamefont {M.~A.}\ \bibnamefont
  {Srokosz}},\ }\bibfield  {title} {\bibinfo {title} {Plankton distribution
  spectra: inter-size class variability and the relative slopes for
  phytoplankton and zooplankton},\ }\href@noop {} {\bibfield  {journal}
  {\bibinfo  {journal} {Geophys. Res. Lett.}\ }\textbf {\bibinfo {volume}
  {29}},\ \bibinfo {pages} {2213–} (\bibinfo {year} {2002})}\BibitemShut
  {NoStop}%
\bibitem [{\citenamefont {L{\'e}vy}\ and\ \citenamefont
  {Klein}(2004)}]{LK2004}%
  \BibitemOpen
  \bibfield  {author} {\bibinfo {author} {\bibfnamefont {M.}~\bibnamefont
  {L{\'e}vy}}\ and\ \bibinfo {author} {\bibfnamefont {P.}~\bibnamefont
  {Klein}},\ }\bibfield  {title} {\bibinfo {title} {Does the low frequency
  variability of mesoscale dynamics explain a part of the phytoplankton and
  zooplankton spectral variability?},\ }\href@noop {} {\bibfield  {journal}
  {\bibinfo  {journal} {Proc. R. Soc. Lond. A}\ }\textbf {\bibinfo {volume}
  {460}},\ \bibinfo {pages} {1673} (\bibinfo {year} {2004})}\BibitemShut
  {NoStop}%
\bibitem [{\citenamefont {Franks}(2005)}]{franks}%
  \BibitemOpen
  \bibfield  {author} {\bibinfo {author} {\bibfnamefont {P.}~\bibnamefont
  {Franks}},\ }\bibfield  {title} {\bibinfo {title} {Plankton patchiness,
  turbulent transport and spatial spectra},\ }\href@noop {} {\bibfield
  {journal} {\bibinfo  {journal} {Marine Ecology Progress Series}\ }\textbf
  {\bibinfo {volume} {294}},\ \bibinfo {pages} {295} (\bibinfo {year}
  {2005})}\BibitemShut {NoStop}%
\bibitem [{\citenamefont {Powell}\ and\ \citenamefont {Okubo}(1994)}]{powell}%
  \BibitemOpen
  \bibfield  {author} {\bibinfo {author} {\bibfnamefont {T.}~\bibnamefont
  {Powell}}\ and\ \bibinfo {author} {\bibfnamefont {A.}~\bibnamefont {Okubo}},\
  }\bibfield  {title} {\bibinfo {title} {Turbulence, diffusion and patchiness
  in the sea},\ }\href@noop {} {\bibfield  {journal} {\bibinfo  {journal}
  {Philosophical Transactions of the Royal Society of London. Series B:
  Biological Sciences}\ }\textbf {\bibinfo {volume} {343}},\ \bibinfo {pages}
  {11} (\bibinfo {year} {1994})}\BibitemShut {NoStop}%
\bibitem [{\citenamefont {Pasquero}\ \emph {et~al.}(2004)\citenamefont
  {Pasquero}, \citenamefont {Bracco},\ and\ \citenamefont
  {Provenzale}}]{pasquero2004coherent}%
  \BibitemOpen
  \bibfield  {author} {\bibinfo {author} {\bibfnamefont {C.}~\bibnamefont
  {Pasquero}}, \bibinfo {author} {\bibfnamefont {A.}~\bibnamefont {Bracco}},\
  and\ \bibinfo {author} {\bibfnamefont {A.}~\bibnamefont {Provenzale}},\
  }\bibfield  {title} {\bibinfo {title} {Coherent vortices, lagrangian
  particles and the marine ecosystem},\ }\href@noop {} {\bibfield  {journal}
  {\bibinfo  {journal} {Shallow flows}\ ,\ \bibinfo {pages} {399}} (\bibinfo
  {year} {2004})}\BibitemShut {NoStop}%
\bibitem [{\citenamefont {Sandulescu}\ \emph {et~al.}(2007)\citenamefont
  {Sandulescu}, \citenamefont {L{\'o}pez}, \citenamefont
  {Hern{\'a}ndez-Garc{\'\i}a},\ and\ \citenamefont
  {Feudel}}]{Sandulescu_etal2007}%
  \BibitemOpen
  \bibfield  {author} {\bibinfo {author} {\bibfnamefont {M.}~\bibnamefont
  {Sandulescu}}, \bibinfo {author} {\bibfnamefont {C.}~\bibnamefont
  {L{\'o}pez}}, \bibinfo {author} {\bibfnamefont {E.}~\bibnamefont
  {Hern{\'a}ndez-Garc{\'\i}a}},\ and\ \bibinfo {author} {\bibfnamefont
  {U.}~\bibnamefont {Feudel}},\ }\bibfield  {title} {\bibinfo {title} {Plankton
  blooms in vortices: the role of biological and hydrodynamic timescales},\
  }\href@noop {} {\bibfield  {journal} {\bibinfo  {journal} {Nonlinear
  Processes in Geophysics}\ }\textbf {\bibinfo {volume} {14}},\ \bibinfo
  {pages} {443} (\bibinfo {year} {2007})}\BibitemShut {NoStop}%
\bibitem [{\citenamefont {Sandulescu}\ \emph {et~al.}(2008)\citenamefont
  {Sandulescu}, \citenamefont {L{\'o}pez}, \citenamefont
  {Hern{\'a}ndez-Garc{\'\i}a},\ and\ \citenamefont
  {Feudel}}]{Sandulescu_etal2008}%
  \BibitemOpen
  \bibfield  {author} {\bibinfo {author} {\bibfnamefont {M.}~\bibnamefont
  {Sandulescu}}, \bibinfo {author} {\bibfnamefont {C.}~\bibnamefont
  {L{\'o}pez}}, \bibinfo {author} {\bibfnamefont {E.}~\bibnamefont
  {Hern{\'a}ndez-Garc{\'\i}a}},\ and\ \bibinfo {author} {\bibfnamefont
  {U.}~\bibnamefont {Feudel}},\ }\bibfield  {title} {\bibinfo {title}
  {Biological activity in the wake of an island close to a coastal upwelling},\
  }\href@noop {} {\bibfield  {journal} {\bibinfo  {journal} {Ecological
  Complexity}\ }\textbf {\bibinfo {volume} {5}},\ \bibinfo {pages} {228}
  (\bibinfo {year} {2008})}\BibitemShut {NoStop}%
\bibitem [{\citenamefont {Hern\'andez-Carrasco}\ \emph
  {et~al.}(2014)\citenamefont {Hern\'andez-Carrasco}, \citenamefont {Rossi},
  \citenamefont {Hern\'andez-Garc\'ia}, \citenamefont {Gar\c{c}on},\ and\
  \citenamefont {L\'opez}}]{Carrasco_etal_2014}%
  \BibitemOpen
  \bibfield  {author} {\bibinfo {author} {\bibfnamefont {I.}~\bibnamefont
  {Hern\'andez-Carrasco}}, \bibinfo {author} {\bibfnamefont {V.}~\bibnamefont
  {Rossi}}, \bibinfo {author} {\bibfnamefont {E.}~\bibnamefont
  {Hern\'andez-Garc\'ia}}, \bibinfo {author} {\bibfnamefont {V.}~\bibnamefont
  {Gar\c{c}on}},\ and\ \bibinfo {author} {\bibfnamefont {C.}~\bibnamefont
  {L\'opez}},\ }\bibfield  {title} {\bibinfo {title} {The reduction of plankton
  biomass induced by mesoscale stirring: A modeling study in the {B}enguela
  upwelling},\ }\href@noop {} {\bibfield  {journal} {\bibinfo  {journal}
  {Deep-Sea Res. I}\ }\textbf {\bibinfo {volume} {83}},\ \bibinfo {pages} {65}
  (\bibinfo {year} {2014})}\BibitemShut {NoStop}%
\bibitem [{\citenamefont {Rossi}\ \emph {et~al.}(2008)\citenamefont {Rossi},
  \citenamefont {L{\'o}pez}, \citenamefont {Sudre}, \citenamefont
  {Hern{\'a}ndez-Garc{\'\i}a},\ and\ \citenamefont
  {Gar{\c{c}}on}}]{rossi2008comparative}%
  \BibitemOpen
  \bibfield  {author} {\bibinfo {author} {\bibfnamefont {V.}~\bibnamefont
  {Rossi}}, \bibinfo {author} {\bibfnamefont {C.}~\bibnamefont {L{\'o}pez}},
  \bibinfo {author} {\bibfnamefont {J.}~\bibnamefont {Sudre}}, \bibinfo
  {author} {\bibfnamefont {E.}~\bibnamefont {Hern{\'a}ndez-Garc{\'\i}a}},\ and\
  \bibinfo {author} {\bibfnamefont {V.}~\bibnamefont {Gar{\c{c}}on}},\
  }\bibfield  {title} {\bibinfo {title} {Comparative study of mixing and
  biological activity of the {B}enguela and {C}anary upwelling systems},\
  }\href@noop {} {\bibfield  {journal} {\bibinfo  {journal} {Geophysical
  Research Letters}\ }\textbf {\bibinfo {volume} {35}} (\bibinfo {year}
  {2008})}\BibitemShut {NoStop}%
\bibitem [{\citenamefont {Rossi}\ \emph {et~al.}(2009)\citenamefont {Rossi},
  \citenamefont {L{\'o}pez}, \citenamefont {Hern{\'a}ndez-Garc{\'\i}a},
  \citenamefont {Sudre}, \citenamefont {Gar{\c{c}}on},\ and\ \citenamefont
  {Morel}}]{rossi2009surface}%
  \BibitemOpen
  \bibfield  {author} {\bibinfo {author} {\bibfnamefont {V.}~\bibnamefont
  {Rossi}}, \bibinfo {author} {\bibfnamefont {C.}~\bibnamefont {L{\'o}pez}},
  \bibinfo {author} {\bibfnamefont {E.}~\bibnamefont
  {Hern{\'a}ndez-Garc{\'\i}a}}, \bibinfo {author} {\bibfnamefont
  {J.}~\bibnamefont {Sudre}}, \bibinfo {author} {\bibfnamefont
  {V.}~\bibnamefont {Gar{\c{c}}on}},\ and\ \bibinfo {author} {\bibfnamefont
  {Y.}~\bibnamefont {Morel}},\ }\bibfield  {title} {\bibinfo {title} {Surface
  mixing and biological activity in the four {E}astern {B}oundary {U}pwelling
  {S}ystems},\ }\href@noop {} {\bibfield  {journal} {\bibinfo  {journal}
  {Nonlinear Processes in Geophysics}\ }\textbf {\bibinfo {volume} {16}},\
  \bibinfo {pages} {557} (\bibinfo {year} {2009})}\BibitemShut {NoStop}%
\bibitem [{\citenamefont {Toroczkai}\ \emph {et~al.}(1998)\citenamefont
  {Toroczkai}, \citenamefont {K{\'a}rolyi}, \citenamefont {P{\'e}ntek},
  \citenamefont {T{\'e}l},\ and\ \citenamefont
  {Grebogi}}]{toroczkai1998advection}%
  \BibitemOpen
  \bibfield  {author} {\bibinfo {author} {\bibfnamefont {Z.}~\bibnamefont
  {Toroczkai}}, \bibinfo {author} {\bibfnamefont {G.}~\bibnamefont
  {K{\'a}rolyi}}, \bibinfo {author} {\bibfnamefont {{\'A}.}~\bibnamefont
  {P{\'e}ntek}}, \bibinfo {author} {\bibfnamefont {T.}~\bibnamefont
  {T{\'e}l}},\ and\ \bibinfo {author} {\bibfnamefont {C.}~\bibnamefont
  {Grebogi}},\ }\bibfield  {title} {\bibinfo {title} {Advection of active
  particles in open chaotic flows},\ }\href@noop {} {\bibfield  {journal}
  {\bibinfo  {journal} {Physical review letters}\ }\textbf {\bibinfo {volume}
  {80}},\ \bibinfo {pages} {500} (\bibinfo {year} {1998})}\BibitemShut
  {NoStop}%
\bibitem [{\citenamefont {Neufeld}\ \emph {et~al.}(2000)\citenamefont
  {Neufeld}, \citenamefont {L{\'o}pez}, \citenamefont
  {Hern{\'a}ndez-Garc{\'\i}a},\ and\ \citenamefont {T{\'e}l}}]{neufeld}%
  \BibitemOpen
  \bibfield  {author} {\bibinfo {author} {\bibfnamefont {Z.}~\bibnamefont
  {Neufeld}}, \bibinfo {author} {\bibfnamefont {C.}~\bibnamefont {L{\'o}pez}},
  \bibinfo {author} {\bibfnamefont {E.}~\bibnamefont
  {Hern{\'a}ndez-Garc{\'\i}a}},\ and\ \bibinfo {author} {\bibfnamefont
  {T.}~\bibnamefont {T{\'e}l}},\ }\bibfield  {title} {\bibinfo {title}
  {Multifractal structure of chaotically advected chemical fields},\
  }\href@noop {} {\bibfield  {journal} {\bibinfo  {journal} {Physical Review
  E}\ }\textbf {\bibinfo {volume} {61}},\ \bibinfo {pages} {3857} (\bibinfo
  {year} {2000})}\BibitemShut {NoStop}%
\bibitem [{\citenamefont {Neufeld}(2001)}]{neu}%
  \BibitemOpen
  \bibfield  {author} {\bibinfo {author} {\bibfnamefont {Z.}~\bibnamefont
  {Neufeld}},\ }\bibfield  {title} {\bibinfo {title} {Excitable media in a
  chaotic flow},\ }\href@noop {} {\bibfield  {journal} {\bibinfo  {journal}
  {Physical review letters}\ }\textbf {\bibinfo {volume} {87}},\ \bibinfo
  {pages} {108301} (\bibinfo {year} {2001})}\BibitemShut {NoStop}%
\bibitem [{\citenamefont {Neufeld}\ \emph
  {et~al.}(2002{\natexlab{a}})\citenamefont {Neufeld}, \citenamefont
  {Hern\'andez-Garc\'ia},\ and\ \citenamefont {Piro}}]{NLHP2002}%
  \BibitemOpen
  \bibfield  {author} {\bibinfo {author} {\bibfnamefont {Z.}~\bibnamefont
  {Neufeld}, \bibfnamefont {C.~L\'opez}}, \bibinfo {author} {\bibfnamefont
  {E.}~\bibnamefont {Hern\'andez-Garc\'ia}},\ and\ \bibinfo {author}
  {\bibfnamefont {O.}~\bibnamefont {Piro}},\ }\bibfield  {title} {\bibinfo
  {title} {Excitable media in open and closed chaotic flow},\ }\href@noop {}
  {\bibfield  {journal} {\bibinfo  {journal} {Phys. Rev. E}\ }\textbf {\bibinfo
  {volume} {66}},\ \bibinfo {pages} {066208} (\bibinfo {year}
  {2002}{\natexlab{a}})}\BibitemShut {NoStop}%
\bibitem [{\citenamefont {Truscott}\ and\ \citenamefont
  {Brindley}(1994{\natexlab{a}})}]{tru}%
  \BibitemOpen
  \bibfield  {author} {\bibinfo {author} {\bibfnamefont {J.}~\bibnamefont
  {Truscott}}\ and\ \bibinfo {author} {\bibfnamefont {J.}~\bibnamefont
  {Brindley}},\ }\bibfield  {title} {\bibinfo {title} {Ocean plankton
  populations as excitable media},\ }\href@noop {} {\bibfield  {journal}
  {\bibinfo  {journal} {Bulletin of Mathematical Biology}\ }\textbf {\bibinfo
  {volume} {56}},\ \bibinfo {pages} {981} (\bibinfo {year}
  {1994}{\natexlab{a}})}\BibitemShut {NoStop}%
\bibitem [{\citenamefont {Truscott}\ and\ \citenamefont
  {Brindley}(1994{\natexlab{b}})}]{truscott1994equilibria}%
  \BibitemOpen
  \bibfield  {author} {\bibinfo {author} {\bibfnamefont {J.}~\bibnamefont
  {Truscott}}\ and\ \bibinfo {author} {\bibfnamefont {J.}~\bibnamefont
  {Brindley}},\ }\bibfield  {title} {\bibinfo {title} {Equilibria, stability
  and excitability in a general class of plankton population models},\
  }\href@noop {} {\bibfield  {journal} {\bibinfo  {journal} {Philosophical
  Transactions of the Royal Society of London. Series A: Physical and
  Engineering Sciences}\ }\textbf {\bibinfo {volume} {347}},\ \bibinfo {pages}
  {703} (\bibinfo {year} {1994}{\natexlab{b}})}\BibitemShut {NoStop}%
\bibitem [{\citenamefont {Sandulescu}\ \emph {et~al.}(2006)\citenamefont
  {Sandulescu}, \citenamefont {Hern{\'a}ndez-Garc{\'\i}a}, \citenamefont
  {L{\'o}pez},\ and\ \citenamefont {Feudel}}]{sandu2006}%
  \BibitemOpen
  \bibfield  {author} {\bibinfo {author} {\bibfnamefont {M.}~\bibnamefont
  {Sandulescu}}, \bibinfo {author} {\bibfnamefont {E.}~\bibnamefont
  {Hern{\'a}ndez-Garc{\'\i}a}}, \bibinfo {author} {\bibfnamefont
  {C.}~\bibnamefont {L{\'o}pez}},\ and\ \bibinfo {author} {\bibfnamefont
  {U.}~\bibnamefont {Feudel}},\ }\bibfield  {title} {\bibinfo {title}
  {Kinematic studies of transport across an island wake, with application to
  the canary islands},\ }\href@noop {} {\bibfield  {journal} {\bibinfo
  {journal} {Tellus A: Dynamic Meteorology and Oceanography}\ }\textbf
  {\bibinfo {volume} {58}},\ \bibinfo {pages} {605} (\bibinfo {year}
  {2006})}\BibitemShut {NoStop}%
\bibitem [{\citenamefont {Murray}(2002)}]{mu}%
  \BibitemOpen
  \bibfield  {author} {\bibinfo {author} {\bibfnamefont {J.~D.}\ \bibnamefont
  {Murray}},\ }\href@noop {} {\emph {\bibinfo {title} {Mathematical Biology}}}\
  (\bibinfo  {publisher} {Springer-Verlag New York},\ \bibinfo {year}
  {2002})\BibitemShut {NoStop}%
\bibitem [{\citenamefont {Grindrod}(1991)}]{gri}%
  \BibitemOpen
  \bibfield  {author} {\bibinfo {author} {\bibfnamefont {P.}~\bibnamefont
  {Grindrod}},\ }\href@noop {} {\emph {\bibinfo {title} {Patterns and Waves :
  Theory and Applications of Reaction-diffusion Equations}}}\ (\bibinfo
  {publisher} {Oxford University Press},\ \bibinfo {year} {1991})\BibitemShut
  {NoStop}%
\bibitem [{\citenamefont {Holling}(1959)}]{holl}%
  \BibitemOpen
  \bibfield  {author} {\bibinfo {author} {\bibfnamefont {C.~S.}\ \bibnamefont
  {Holling}},\ }\bibfield  {title} {\bibinfo {title} {The components of
  predation as revealed by a study of small-mammal predation of the european
  pine sawfly},\ }\href@noop {} {\bibfield  {journal} {\bibinfo  {journal} {The
  Canadian Entomologist}\ }\textbf {\bibinfo {volume} {91}},\ \bibinfo {pages}
  {293–320} (\bibinfo {year} {1959})}\BibitemShut {NoStop}%
\bibitem [{\citenamefont {Hern{\'a}ndez-Garc{\'\i}a}\ \emph
  {et~al.}(2002)\citenamefont {Hern{\'a}ndez-Garc{\'\i}a}, \citenamefont
  {L{\'o}pez},\ and\ \citenamefont {Neufeld}}]{chaos}%
  \BibitemOpen
  \bibfield  {author} {\bibinfo {author} {\bibfnamefont {E.}~\bibnamefont
  {Hern{\'a}ndez-Garc{\'\i}a}}, \bibinfo {author} {\bibfnamefont
  {C.}~\bibnamefont {L{\'o}pez}},\ and\ \bibinfo {author} {\bibfnamefont
  {Z.}~\bibnamefont {Neufeld}},\ }\bibfield  {title} {\bibinfo {title}
  {Small-scale structure of nonlinearly interacting species advected by chaotic
  flows},\ }\href@noop {} {\bibfield  {journal} {\bibinfo  {journal} {Chaos: An
  Interdisciplinary Journal of Nonlinear Science}\ }\textbf {\bibinfo {volume}
  {12}},\ \bibinfo {pages} {470} (\bibinfo {year} {2002})}\BibitemShut
  {NoStop}%
\bibitem [{\citenamefont {Neufeld}\ \emph
  {et~al.}(2002{\natexlab{b}})\citenamefont {Neufeld}, \citenamefont {Haynes},
  \citenamefont {Gar{\c{c}}on},\ and\ \citenamefont {Sudre}}]{fer}%
  \BibitemOpen
  \bibfield  {author} {\bibinfo {author} {\bibfnamefont {Z.}~\bibnamefont
  {Neufeld}}, \bibinfo {author} {\bibfnamefont {P.}~\bibnamefont {Haynes}},
  \bibinfo {author} {\bibfnamefont {V.}~\bibnamefont {Gar{\c{c}}on}},\ and\
  \bibinfo {author} {\bibfnamefont {J.}~\bibnamefont {Sudre}},\ }\bibfield
  {title} {\bibinfo {title} {Ocean fertilization experiments may initiate a
  large scale phytoplankton bloom},\ }\href@noop {} {\bibfield  {journal}
  {\bibinfo  {journal} {Geophysical research letters}\ }\textbf {\bibinfo
  {volume} {29}},\ \bibinfo {pages} {29} (\bibinfo {year}
  {2002}{\natexlab{b}})}\BibitemShut {NoStop}%
\bibitem [{\citenamefont {Polin}\ \emph {et~al.}(2009)\citenamefont {Polin},
  \citenamefont {Tuval}, \citenamefont {Drescher}, \citenamefont {Gollub},\
  and\ \citenamefont {Goldstein}}]{Polin_etal2009}%
  \BibitemOpen
  \bibfield  {author} {\bibinfo {author} {\bibfnamefont {M.}~\bibnamefont
  {Polin}}, \bibinfo {author} {\bibfnamefont {I.}~\bibnamefont {Tuval}},
  \bibinfo {author} {\bibfnamefont {K.}~\bibnamefont {Drescher}}, \bibinfo
  {author} {\bibfnamefont {J.}~\bibnamefont {Gollub}},\ and\ \bibinfo {author}
  {\bibfnamefont {R.}~\bibnamefont {Goldstein}},\ }\bibfield  {title} {\bibinfo
  {title} {Chlamydomonas swims with two “gears” in a eukaryotic version of
  run-and-tumble locomotion},\ }\href@noop {} {\bibfield  {journal} {\bibinfo
  {journal} {Science}\ }\textbf {\bibinfo {volume} {325}},\ \bibinfo {pages}
  {487} (\bibinfo {year} {2009})}\BibitemShut {NoStop}%
\bibitem [{\citenamefont {Garcia}\ \emph {et~al.}(2011)\citenamefont {Garcia},
  \citenamefont {Berti}, \citenamefont {Peyla},\ and\ \citenamefont
  {Rafa{\"\i}}}]{Garcia_etal2011}%
  \BibitemOpen
  \bibfield  {author} {\bibinfo {author} {\bibfnamefont {M.}~\bibnamefont
  {Garcia}}, \bibinfo {author} {\bibfnamefont {S.}~\bibnamefont {Berti}},
  \bibinfo {author} {\bibfnamefont {P.}~\bibnamefont {Peyla}},\ and\ \bibinfo
  {author} {\bibfnamefont {S.}~\bibnamefont {Rafa{\"\i}}},\ }\bibfield  {title}
  {\bibinfo {title} {Random walk of a swimmer in a low-{R}eynolds-number
  medium},\ }\href@noop {} {\bibfield  {journal} {\bibinfo  {journal} {Physical
  Review E}\ }\textbf {\bibinfo {volume} {83}},\ \bibinfo {pages} {035301(R)}
  (\bibinfo {year} {2011})}\BibitemShut {NoStop}%
\bibitem [{\citenamefont {Brun-Cosme-Bruny}\ \emph {et~al.}(2019)\citenamefont
  {Brun-Cosme-Bruny}, \citenamefont {Bertin}, \citenamefont {Coasne},
  \citenamefont {Peyla},\ and\ \citenamefont {Rafa{\"i}}}]{Brun_etal2019}%
  \BibitemOpen
  \bibfield  {author} {\bibinfo {author} {\bibfnamefont {M.}~\bibnamefont
  {Brun-Cosme-Bruny}}, \bibinfo {author} {\bibfnamefont {E.}~\bibnamefont
  {Bertin}}, \bibinfo {author} {\bibfnamefont {B.}~\bibnamefont {Coasne}},
  \bibinfo {author} {\bibfnamefont {P.}~\bibnamefont {Peyla}},\ and\ \bibinfo
  {author} {\bibfnamefont {S.}~\bibnamefont {Rafa{\"i}}},\ }\bibfield  {title}
  {\bibinfo {title} {Effective diffusivity of micro-swimmers in a crowded
  environment},\ }\href@noop {} {\bibfield  {journal} {\bibinfo  {journal} {The
  Journal of chemical physics}\ }\textbf {\bibinfo {volume} {150}},\ \bibinfo
  {pages} {104901} (\bibinfo {year} {2019})}\BibitemShut {NoStop}%
\bibitem [{\citenamefont {Boffetta}\ and\ \citenamefont {Ecke}(2012)}]{boff}%
  \BibitemOpen
  \bibfield  {author} {\bibinfo {author} {\bibfnamefont {G.}~\bibnamefont
  {Boffetta}}\ and\ \bibinfo {author} {\bibfnamefont {R.}~\bibnamefont
  {Ecke}},\ }\bibfield  {title} {\bibinfo {title} {Two-dimensional
  turbulence},\ }\href@noop {} {\bibfield  {journal} {\bibinfo  {journal}
  {Annual Review of Fluid Mechanics}\ }\textbf {\bibinfo {volume} {44}},\
  \bibinfo {pages} {427} (\bibinfo {year} {2012})}\BibitemShut {NoStop}%
\bibitem [{\citenamefont {Bastine}\ and\ \citenamefont
  {Feudel}(2010)}]{BF2010}%
  \BibitemOpen
  \bibfield  {author} {\bibinfo {author} {\bibfnamefont {D.}~\bibnamefont
  {Bastine}}\ and\ \bibinfo {author} {\bibfnamefont {U.}~\bibnamefont
  {Feudel}},\ }\bibfield  {title} {\bibinfo {title} {Inhomogeneous dominance
  patterns of competing phytoplankton groups in the wake of an island},\
  }\href@noop {} {\bibfield  {journal} {\bibinfo  {journal} {Nonlinear
  Processes in Geophysics}\ }\textbf {\bibinfo {volume} {17}},\ \bibinfo
  {pages} {715} (\bibinfo {year} {2010})}\BibitemShut {NoStop}%
\bibitem [{\citenamefont {Van~Dyke}(1982)}]{vd}%
  \BibitemOpen
  \bibfield  {author} {\bibinfo {author} {\bibfnamefont {M.}~\bibnamefont
  {Van~Dyke}},\ }\href@noop {} {\emph {\bibinfo {title} {An album of fluid
  motion}}}\ (\bibinfo  {publisher} {Parabolic Press Stanford},\ \bibinfo
  {year} {1982})\BibitemShut {NoStop}%
\bibitem [{\citenamefont {Zdravkovich}(1997)}]{z}%
  \BibitemOpen
  \bibfield  {author} {\bibinfo {author} {\bibfnamefont {M.~M.}\ \bibnamefont
  {Zdravkovich}},\ }\href@noop {} {\emph {\bibinfo {title} {Flow Around
  Circular Cylinders Volume 1: Fundamentals}}}\ (\bibinfo  {publisher} {Oxford
  University Press},\ \bibinfo {year} {1997})\BibitemShut {NoStop}%
\bibitem [{\citenamefont {Kraichnan}(1967)}]{kra}%
  \BibitemOpen
  \bibfield  {author} {\bibinfo {author} {\bibfnamefont {R.}~\bibnamefont
  {Kraichnan}},\ }\bibfield  {title} {\bibinfo {title} {Inertial ranges in
  two-dimensional turbulence},\ }\href@noop {} {\bibfield  {journal} {\bibinfo
  {journal} {The Physics of Fluids}\ }\textbf {\bibinfo {volume} {10}},\
  \bibinfo {pages} {1417} (\bibinfo {year} {1967})}\BibitemShut {NoStop}%
\bibitem [{\citenamefont {Batchelor}(1959)}]{batchelor1959small}%
  \BibitemOpen
  \bibfield  {author} {\bibinfo {author} {\bibfnamefont {G.~K.}\ \bibnamefont
  {Batchelor}},\ }\bibfield  {title} {\bibinfo {title} {Small-scale variation
  of convected quantities like temperature in turbulent fluid part 1. general
  discussion and the case of small conductivity},\ }\href@noop {} {\bibfield
  {journal} {\bibinfo  {journal} {Journal of Fluid Mechanics}\ }\textbf
  {\bibinfo {volume} {5}},\ \bibinfo {pages} {113} (\bibinfo {year}
  {1959})}\BibitemShut {NoStop}%
\bibitem [{\citenamefont {Ruelle}(1979)}]{ruelle}%
  \BibitemOpen
  \bibfield  {author} {\bibinfo {author} {\bibfnamefont {D.}~\bibnamefont
  {Ruelle}},\ }\bibfield  {title} {\bibinfo {title} {Microscopic fluctuations
  and turbulence},\ }\href@noop {} {\bibfield  {journal} {\bibinfo  {journal}
  {Physics Letters A}\ }\textbf {\bibinfo {volume} {72}},\ \bibinfo {pages}
  {81} (\bibinfo {year} {1979})}\BibitemShut {NoStop}%
\bibitem [{\citenamefont {Hern{\'a}ndez-Carrasco}\ \emph
  {et~al.}(2012)\citenamefont {Hern{\'a}ndez-Carrasco}, \citenamefont
  {L{\'o}pez}, \citenamefont {Hern{\'a}ndez-Garc{\'\i}a},\ and\ \citenamefont
  {Turiel}}]{hernandez2012seasonal}%
  \BibitemOpen
  \bibfield  {author} {\bibinfo {author} {\bibfnamefont {I.}~\bibnamefont
  {Hern{\'a}ndez-Carrasco}}, \bibinfo {author} {\bibfnamefont {C.}~\bibnamefont
  {L{\'o}pez}}, \bibinfo {author} {\bibfnamefont {E.}~\bibnamefont
  {Hern{\'a}ndez-Garc{\'\i}a}},\ and\ \bibinfo {author} {\bibfnamefont
  {A.}~\bibnamefont {Turiel}},\ }\bibfield  {title} {\bibinfo {title} {Seasonal
  and regional characterization of horizontal stirring in the global ocean},\
  }\href@noop {} {\bibfield  {journal} {\bibinfo  {journal} {Journal of
  Geophysical Research: Oceans}\ }\textbf {\bibinfo {volume} {117}} (\bibinfo
  {year} {2012})}\BibitemShut {NoStop}%

\bibitem [{\citenamefont {Mandelbrot}(1967)}]{Mandelbrot1967}%
  \BibitemOpen
  \bibfield  {author} {\bibinfo {author} {\bibfnamefont {B.}~\bibnamefont
  {Mandelbrot}},\ }\bibfield  {title} {\bibinfo {title} {How long is the coast
  of {B}ritain? statistical self-similarity and fractional dimension},\
  }\href@noop {} {\bibfield  {journal} {\bibinfo  {journal} {Science}\ }\textbf
  {\bibinfo {volume} {156}},\ \bibinfo {pages} {636} (\bibinfo {year}
  {1967})}\BibitemShut {NoStop}%
\bibitem [{\citenamefont {Boffetta}\ \emph {et~al.}(2008)\citenamefont
  {Boffetta}, \citenamefont {Celani}, \citenamefont {Dezzani},\ and\
  \citenamefont {Seminara}}]{BCDS2008}%
  \BibitemOpen
  \bibfield  {author} {\bibinfo {author} {\bibfnamefont {G.}~\bibnamefont
  {Boffetta}}, \bibinfo {author} {\bibfnamefont {A.}~\bibnamefont {Celani}},
  \bibinfo {author} {\bibfnamefont {D.}~\bibnamefont {Dezzani}},\ and\ \bibinfo
  {author} {\bibfnamefont {A.}~\bibnamefont {Seminara}},\ }\bibfield  {title}
  {\bibinfo {title} {How winding is the coast of {B}ritain? conformal
  invariance of rocky shorelines},\ }\href@noop {} {\bibfield  {journal}
  {\bibinfo  {journal} {Geophysical research letters}\ }\textbf {\bibinfo
  {volume} {35}} (\bibinfo {year} {2008})}\BibitemShut {NoStop}%
\bibitem [{\citenamefont {Toppaladoddi}\ \emph {et~al.}(2020)\citenamefont
  {Toppaladoddi}, \citenamefont {Wells}, \citenamefont {Doering},\ and\
  \citenamefont {Wettlaufer}}]{rough}%
  \BibitemOpen
  \bibfield  {author} {\bibinfo {author} {\bibfnamefont {S.}~\bibnamefont
  {Toppaladoddi}}, \bibinfo {author} {\bibfnamefont {A.~J.}\ \bibnamefont
  {Wells}}, \bibinfo {author} {\bibfnamefont {C.}~\bibnamefont {Doering}},\
  and\ \bibinfo {author} {\bibfnamefont {J.}~\bibnamefont {Wettlaufer}},\
  }\bibfield  {title} {\bibinfo {title} {Thermal convection over fractal
  surfaces},\ }\href@noop {} {\bibfield  {journal} {\bibinfo  {journal}
  {Journal of Fluid Mechanics}\ }\textbf {\bibinfo {volume} {907}} (\bibinfo
  {year} {2020})}\BibitemShut {NoStop}%
\bibitem [{\citenamefont {Pope}(2001)}]{pope}%
  \BibitemOpen
  \bibfield  {author} {\bibinfo {author} {\bibfnamefont {S.~B.}\ \bibnamefont
  {Pope}},\ }\href@noop {} {\bibinfo {title} {Turbulent flows}} (\bibinfo
  {year} {2001})\BibitemShut {NoStop}%
\bibitem [{\citenamefont {Martin}(2000)}]{martin2000filament}%
  \BibitemOpen
  \bibfield  {author} {\bibinfo {author} {\bibfnamefont {A.}~\bibnamefont
  {Martin}},\ }\bibfield  {title} {\bibinfo {title} {On filament width in
  oceanic plankton distributions},\ }\href@noop {} {\bibfield  {journal}
  {\bibinfo  {journal} {Journal of plankton research}\ }\textbf {\bibinfo
  {volume} {22}},\ \bibinfo {pages} {597} (\bibinfo {year} {2000})}\BibitemShut
  {NoStop}%
  \bibitem [{\citenamefont {van Hooft}\ \emph {et~al.}(2018)\citenamefont {van
  Hooft}, \citenamefont {Popinet}, \citenamefont {van Heerwaarden},
  \citenamefont {van~der Linden}, \citenamefont {de~Roode},\ and\ \citenamefont
  {van~de Wiel}}]{van}%
  \BibitemOpen
  \bibfield  {author} {\bibinfo {author} {\bibfnamefont {J.}~\bibnamefont {van
  Hooft}}, \bibinfo {author} {\bibfnamefont {S.}~\bibnamefont {Popinet}},
  \bibinfo {author} {\bibfnamefont {C.}~\bibnamefont {van Heerwaarden}},
  \bibinfo {author} {\bibfnamefont {S.}~\bibnamefont {van~der Linden}},
  \bibinfo {author} {\bibfnamefont {S.}~\bibnamefont {de~Roode}},\ and\
  \bibinfo {author} {\bibfnamefont {B.}~\bibnamefont {van~de Wiel}},\
  }\bibfield  {title} {\bibinfo {title} {Towards adaptive grids for atmospheric
  boundary-layer simulations},\ }\href@noop {} {\bibfield  {journal} {\bibinfo
  {journal} {Boundary-layer meteorology}\ }\textbf {\bibinfo {volume} {167}},\
  \bibinfo {pages} {421} (\bibinfo {year} {2018})}\BibitemShut {NoStop}%
   \bibitem [{\citenamefont {Johansen}\ \emph {et~al.}(1998)\citenamefont { ohansen}, 
  \citenamefont {Colella}}]{Johansen1998ACG}%
  \BibitemOpen
  \bibfield  {author} {\bibinfo {author} {\bibfnamefont {H.}~\bibnamefont {Johansen}}, \bibinfo {author} {\bibfnamefont {P.}~\bibnamefont {Colella}},\
  }\bibfield  {title} {\bibinfo {title} { Cartesian Grid Embedded Boundary Method for Poisson's Equation on Irregular Domains},\ }\href@noop {} {\bibfield  {journal} {\bibinfo
  {journal} {Journal of Computational Physics}\ }\textbf {\bibinfo {volume} {147}},\
  \bibinfo {pages} {60-85} (\bibinfo {year} {1998})}\BibitemShut {NoStop}%
\bibitem [{\citenamefont {Popinet}(2009)}]{pop2009}%
  \BibitemOpen
  \bibfield  {author} {\bibinfo {author} {\bibfnamefont {S.}~\bibnamefont
  {Popinet}},\ }\bibfield  {title} {\bibinfo {title} {An accurate adaptive
  solver for surface-tension-driven interfacial flows},\ }\href@noop {}
  {\bibfield  {journal} {\bibinfo  {journal} {Journal of Computational
  Physics}\ }\textbf {\bibinfo {volume} {228}},\ \bibinfo {pages} {5838}
  (\bibinfo {year} {2009})}\BibitemShut {NoStop}%
\bibitem [{\citenamefont {Chorin}(1969)}]{cho}%
  \BibitemOpen
  \bibfield  {author} {\bibinfo {author} {\bibfnamefont {A.}~\bibnamefont
  {Chorin}},\ }\bibfield  {title} {\bibinfo {title} {On the convergence of
  discrete approximations to the {N}avier-{S}tokes equations},\ }\href@noop {}
  {\bibfield  {journal} {\bibinfo  {journal} {Mathematics of computation}\
  }\textbf {\bibinfo {volume} {23}},\ \bibinfo {pages} {341} (\bibinfo {year}
  {1969})}\BibitemShut {NoStop}%
\bibitem [{\citenamefont {Popinet}(2003)}]{pop2003}%
  \BibitemOpen
  \bibfield  {author} {\bibinfo {author} {\bibfnamefont {S.}~\bibnamefont
  {Popinet}},\ }\bibfield  {title} {\bibinfo {title} {Gerris: a tree-based
  adaptive solver for the incompressible {E}uler equations in complex
  geometries},\ }\href@noop {} {\bibfield  {journal} {\bibinfo  {journal}
  {Journal of Computational Physics}\ }\textbf {\bibinfo {volume} {190}},\
  \bibinfo {pages} {572} (\bibinfo {year} {2003})}\BibitemShut {NoStop}%
\bibitem [{\citenamefont {Lagr{\'e}e}\ \emph {et~al.}(2011)\citenamefont
  {Lagr{\'e}e}, \citenamefont {Staron},\ and\ \citenamefont {Popinet}}]{lag}%
  \BibitemOpen
  \bibfield  {author} {\bibinfo {author} {\bibfnamefont {P.}~\bibnamefont
  {Lagr{\'e}e}}, \bibinfo {author} {\bibfnamefont {L.}~\bibnamefont {Staron}},\
  and\ \bibinfo {author} {\bibfnamefont {S.}~\bibnamefont {Popinet}},\
  }\bibfield  {title} {\bibinfo {title} {The granular column collapse as a
  continuum: validity of a two-dimensional {N}avier-{S}tokes model with a
  [mu](i)-rheology},\ }\href@noop {} {\bibfield  {journal} {\bibinfo  {journal}
  {Journal of Fluid Mechanics}\ }\textbf {\bibinfo {volume} {686}},\ \bibinfo
  {pages} {378} (\bibinfo {year} {2011})}\BibitemShut {NoStop}%
\bibitem [{\citenamefont {Popinet}(2015)}]{pop2015}%
  \BibitemOpen
  \bibfield  {author} {\bibinfo {author} {\bibfnamefont {S.}~\bibnamefont
  {Popinet}},\ }\bibfield  {title} {\bibinfo {title} {A quadtree-adaptive
  multigrid solver for the {S}erre--{G}reen--{N}aghdi equations},\ }\href@noop
  {} {\bibfield  {journal} {\bibinfo  {journal} {Journal of Computational
  Physics}\ }\textbf {\bibinfo {volume} {302}},\ \bibinfo {pages} {336}
  (\bibinfo {year} {2015})}\BibitemShut {NoStop}%
 

  
\end{thebibliography}
%

\end{document}